\documentclass{cimento}
\usepackage[dvips]{graphicx}
\usepackage[latin1]{inputenc}
\usepackage{xcolor,bm}
\usepackage{nicefrac}
\usepackage{rotating}
\usepackage{tabularx}

\title{Iron-based superconductors: tales from the nuclei}
\author{Pietro Carretta \from{ins:x} \atque Giacomo Prando \from{ins:x}}
\instlist{\inst{ins:x} Department of Physics, University of Pavia, I-27100 Pavia, Italy}

\begin{document}

\maketitle

\begin{abstract}
High-temperature superconductivity in Fe-based pnictides and chalcogenides has been one of the most significant recent discoveries in condensed matter physics and has attracted remarkable attention in the last decade. These materials are characterized by a complex fermiology and, as a result, feature a wide range of electronic properties as a function of different tuning parameters such as chemical doping, temperature and pressure. Along the path towards the comprehension of the physical mechanisms underlying this rich phenomenology, NMR (nuclear magnetic resonance) and NQR (nuclear quadrupole resonance) have played a role of capital importance that we review in this work. In particular, we address how NMR has contributed to the current understanding of the main regions of the electronic phase diagram of Fe-based pnictides, that is, the -- sometimes coexisting -- antiferromagnetic spin-density wave and superconducting states. We evidence the unique capability of NMR as local-probe technique of investigating the effect of quenched disorder and chemical impurities. Then, we review the NMR signatures of low-frequency fluctuations associated with the development of electronic nematicity as well as with the motion of superconducting flux lines. Finally, we discuss recent contributions of NMR and NQR which evidence an intrinsically inhomogeneous electronic charge distribution as well as an orbitally-selective behaviour. 
\end{abstract}

\tableofcontents

\section{Introduction to iron-based superconductors}

It is said that rules are made to be broken, and often the implications of this attitude for transgression have been highly beneficial for the advancement of Science. A good example comes from the research on superconductivity, a thermodynamical phase of matter characterized by perfect diamagnetism and infinite electrical conductivity. The value of the critical temperature $T_{\rm{c}}$ -- separating superconductivity from the high-temperature conventional (or ``normal'') state -- is the most immediate quantifier of the potential of the considered superconductor for technological applications. Accordingly, the discovery of materials characterized by higher and higher $T_{\rm{c}}$ values has been the main goal of the fundamental research on superconductivity since its discovery by Heike Kamerlingh Onnes in 1911 \cite{Del10}. The empirical rules formulated in the 1950's by Bernd Theodor Matthias, cleverly extrapolating from the common trends observed within known superconducting alloys and binary compounds, played a role of cardinal importance in this sense \cite{Maz10}. In fact, thanks to those rules, the record value $T_{\rm{c}} \simeq 18$ K for the A15 intermetallic Nb$_{3}$Sn reported in 1954 could be improved to $23$ K for Nb$_{3}$Ge. Still, the realization of such improvement took some 20 years, supporting a widespread belief that a saturation value for $T_{\rm{c}}$ had been reached. It was only after 1986 that the discovery of superconductivity in cuprates with $T_{\rm{c}}$ up to about $140$ K changed the situation dramatically, and this turning point was made possible by following strategies almost at variance with the rules put forward by Matthias \cite{Bed86,Lee06,Kei15}. 

The discovery of high-$T_{\rm{c}}$ superconductivity in iron-based pnictides and chalcogenides also marked a clear departure from the Matthias's rules. Most remarkably, the very presence of stoichiometric amounts of paramagnetic Fe within the considered materials could be deemed as detrimental to the development of superconductivity just in view of the lapidary prescription ``stay away from magnetism''. For this reason, the report in 2006 of $T_{\rm{c}} \simeq 4$ K in LaFePO$_{1-x}$F$_{x}$ \cite{Kam06} and in 2008 of $T_{\rm{c}} \simeq 26$ K in LaFeAsO$_{1-x}$F$_{x}$ \cite{Kam08} was received with surprise within the scientific community. The worldwide intense research activity that followed those reports has shown not only that the phenomenon is robustly observed in different classes of materials, but that $T_{\rm{c}}$ values as high as 55 K in bulk samples \cite{Ren08} and 100 K in ultra-thin films \cite{Ge15} can be achieved.

Several review articles have been published over the last decade discussing the physics of IBSs (iron-based superconductors) from a broad perspective \cite{Joh10,Pag10,Ste11,Wen11,Don13,Che14,Mar16,Si16}. In what follows, we summarize the main properties of these materials with particular emphasis to those aspects relevant for our subsequent discussion on some of the most recent contributions of nuclear magnetic resonance to this field of research.

\subsection{Structural properties} 
\begin{figure}
	\vspace{7.5cm} \includegraphics{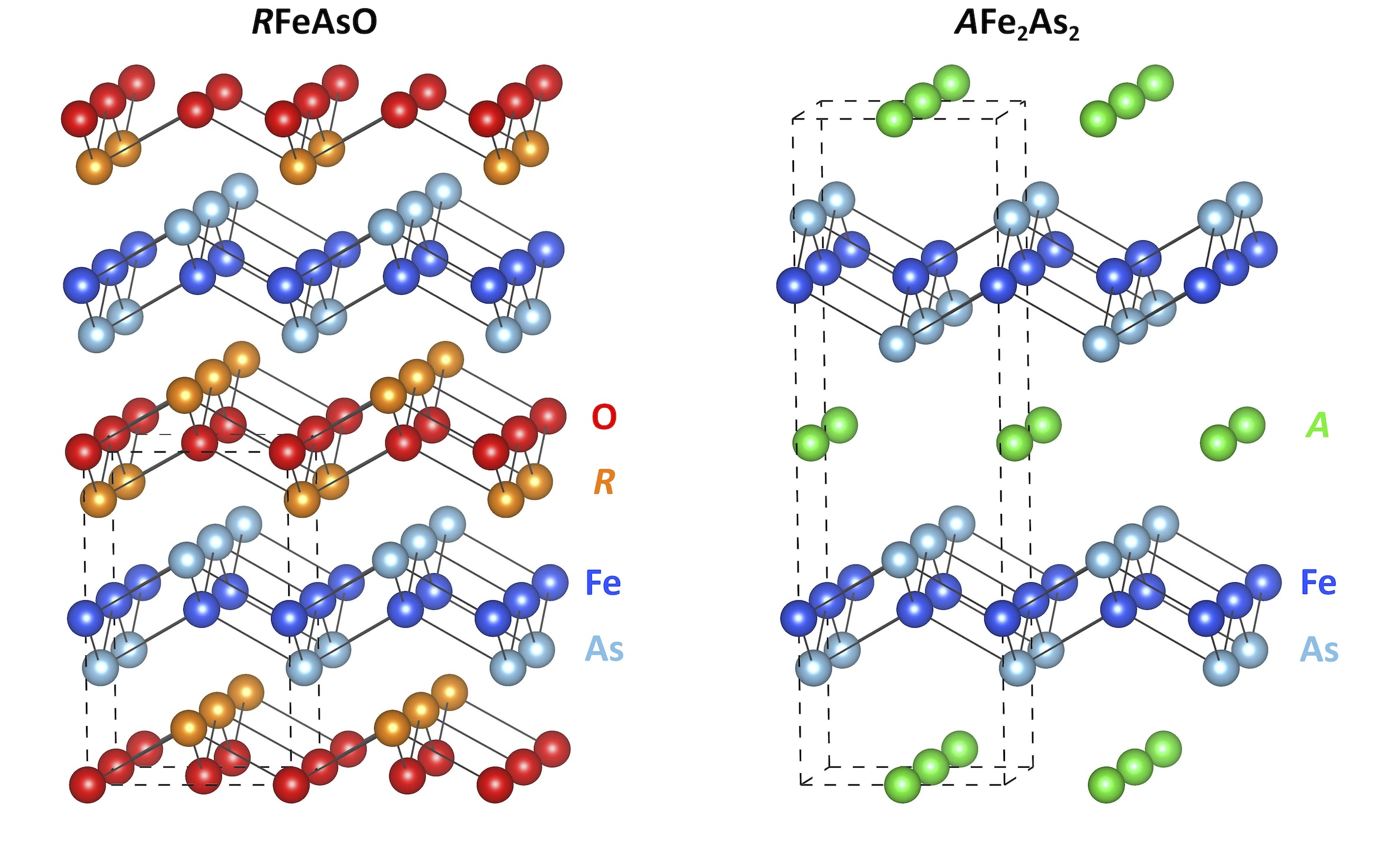} \caption{Representative crystallographic structures for the high-temperature tetragonal phases of iron-based 1111 (left-hand panel, space group \textit{P}4/\textit{nmm}) and 122 (right-hand panel, space group \textit{I}4/\textit{mmm}) pnictides. The dashed lines delimit the unit cell, which contains two formula units for both 1111 and 122 materials. Images generated with the software VESTA \cite{Mom11}.}\label{FigStructure}
\end{figure}

To date, all the known classes of iron-based materials displaying high-$T_{\rm{c}}$ superconductivity share common structural properties. In particular, the fundamental crystallographic units are bi-dimensional arrays of Fe ions arranged at the vertices of a square lattice and sandwiched between two similar layers composed of a pnictogen or chalcogen element -- typically As or Se, respectively. Every Fe ion is positioned in the centre of a distorted tetrahedron with As or Se ions at its vertices. The resulting anisotropic structure is highly reminiscent of cuprates (see Fig.~\ref{FigStructure}).

The simplest structure is observed in the so-called 11 family of chalcogenides -- with FeSe as representative material, whose relative stoichiometric weights justify the name of the family -- where the Fe-Se layers described above are simply stacked one on the top of the other \cite{Hsu08}. For the 1111 pnictides, such as LaFeAsO, the situation is identical to 11's except that a full substitution of O for Fe and of $R$ for As, $R$ being a rare-earth element, takes place in every second layer (see the left-hand panel of Fig.~\ref{FigStructure}). Higher degrees of complexity are obtained by introducing inter-layer structural objects between the layers. One possible strategy is that of intercalating known structures with complex molecules, as suggested for FeSe \cite{Bur13}. Alternatively, these spacing elements can be characterized by long-ranged structural order across the sample. This is the case for the materials isostructural to $A$Fe$_{2}$As$_{2}$, belonging to the 122 family, where Fe-As layers are separated by monolayers composed of alkaline earth ions $A$ arranged in a square lattice (see the right-hand panel of Fig.~\ref{FigStructure}) \cite{Rot08} and for those isostructural to $B$FeAs, belonging to the 111 family, where the spacing elements are two stacked square lattices composed of alkali metals $B$ \cite{Tap08}. Finally, in the more complicated structures reported for, e.g., Sr$_{4}$V$_{2}$O$_{6}$Fe$_{2}$As$_{2}$ -- the representative material of the 42622 family -- the crystal can be thought as a natural stack of Fe-As layers, Sr monolayers and SrVO$_{3}$ elements, the latter with composition and structure resembling perovskite oxides \cite{Ogi09,Zhu09}.

More realistic approaches must be followed in order to go beyond the idealized properties of the lattices described above. For the specific example of alkaline iron selenides belonging to the 122 family, the chemical formula $A_{x}$Fe$_{2-y}$Se$_{2}$ should be used in order to take the atomic vacancies at the $A$ and Fe sites into account. In fact, interesting structural effects are associated with these defects. In particular, it is known that vacancies on the Fe layer tend to give rise to ordered superstructures which are long-ranged for specific compositions \cite{Dag13}. This is the case for, e.g., $A$Fe$_{1.5}$Se$_{2}$ and $A$Fe$_{1.6}$Se$_{2}$ where the ordering of Fe vacancies develops in bidimensional arrangements \cite{Fan11}. This tendency is particularly marked in $A$Fe$_{2}$Se$_{3}$, where the planar squared arrangement of Fe ions described above is cut into quasi-1D (one-dimensional) two-leg ladders by stripe-ordered Fe vacancies \cite{Car11,Lei11,Sap11,Tak15}.

In most iron-based pnictides and chalcogenides, a structural distortion develops at a critical temperature $T_{\rm{s}}$. At high temperatures, compounds belonging to the 1111, 11 and 122 families belong to tetragonal space groups (primitive \textit{P}4/\textit{nmm} and body-centred \textit{I}4/\textit{mmm} for 1111/11 and 122, respectively) but the symmetry is lowered to orthorhombic for $T < T_{\rm{s}}$ (base-centred $Cmme$ and face-centred $Fmmm$ for 1111/11 and 122, respectively), making the two in-plane lattice parameters inequivalent ($b/a > 0.99$). As discussed later on in detail, major effects are reported for the anisotropy of the electronic properties below $T_{\rm{s}}$ in spite of the quantitatively small effect on the structure.
\begin{figure}
	\vspace{5.9cm} \includegraphics{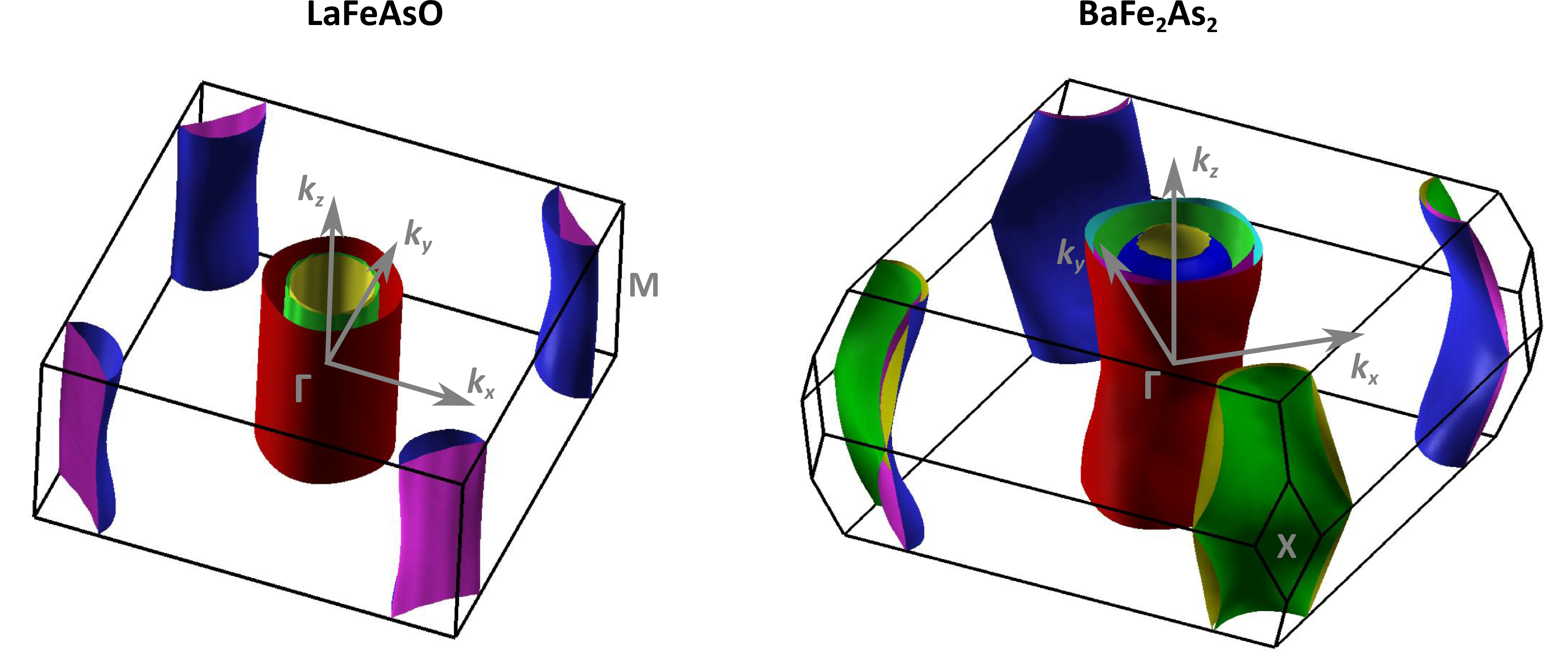} \caption{Fermi surfaces for LaFeAsO (left-hand panel) and BaFe$_{2}$As$_{2}$ (right-hand panel) calculated by means of density functional theory (full-potential linearized augmented plane wave method with generalized gradient approximation) implemented with the Elk code \cite{wwwElk}. The labels of significant high-symmetry points are reported for clarity. Images adapted from \cite{Mor17}.}\label{FigFS}
\end{figure}

\subsection{Electronic band structure and magnetism} The physical properties of pnictide compounds belonging to the 1111 and 122 families are generally modelled by considering the Fe-As layers as the electronically-active elements. Experiments and band-structure calculations evidence that the Fermi level is crossed by the five bands originating from the $d$-orbitals from Fe, with minor contributions to the density of states from those associated with As orbitals \cite{Sin08}. In view of the marked structural anisotropy, the band dispersion along the $k_{z}$ axis in the reciprocal space is not as marked as in the $k_{x}$-$k_{y}$ directions. The resulting Fermi surface is composed of warped cylindrical-like sheets centred at different regions of the BZ (Brillouin zone). As shown in Fig.~\ref{FigFS}, different concentric sheets are located at the centre (corners) of the BZ and enclose hole-like (electron-like) states in so-called ``pockets''. The degree of warping is higher for BaFe$_{2}$As$_{2}$ than for LaFeAsO, suggesting higher bi-dimensionality -- i.e., weaker inter-layer coupling -- in the latter material. Interestingly, wavevectors connecting the centre to the corners of the BZ approximately behave as nesting vectors for the Fermi surface. Overall, this fermiology results in a semimetallic behaviour in the normal state, in strong contrast with the Mott-insulating state observed in the parent compositions of cuprates \cite{Lee06,Kei15}.

\begin{figure}
	\vspace{7.2cm} \includegraphics{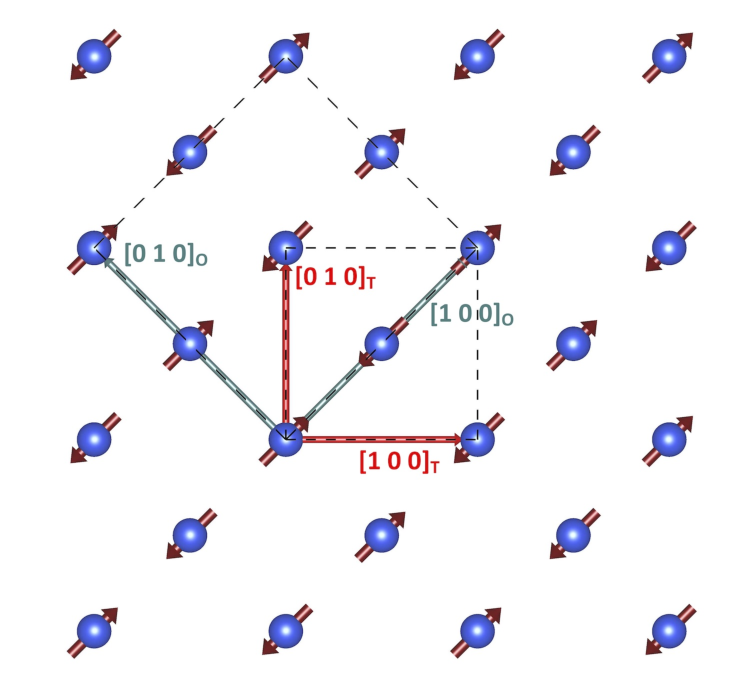} \caption{Microscopic stripy magnetic arrangement on the Fe layer within the SDW phase. The unit vectors of the two in-plane axes have been added for both the tetragonal and the orthorhombic notations (labelled as T and O, respectively). The dashed lines represent the in-plane unit cells in the two notations. Image generated with the software VESTA \cite{Mom11}.}\label{FigAFM}
\end{figure}

Experiments evidence that a commensurate antiferromagnetic SDW (spin-density wave) phase develops below the critical temperature $T_{\rm{N}} \sim 140$ K in 1111 and 122 pnictide materials \cite{Lum10,Dai15}. Remarkably, the magnetic phase develops only within the orthorhombic phase -- i.e., $T_{\rm{N}} \leq T_{\rm{s}}$ -- and the ordered magnetic moment lies in-plane, aligned along the longer orthorhombic axis. The microscopic arrangement of magnetic moments is stripy-like with propagation vector $\left(\frac{\pi}{a},\frac{\pi}{a}\right)$ in tetragonal notation or, equivalently, $\left(\frac{\pi}{a},0\right)$ in the more commonly used orthorhombic notation -- see Fig.~\ref{FigAFM}. The close analogy with the nesting vector strongly supports the view that the development of the SDW phase is driven by the nesting of the Fermi surface. In further overall agreement with the itinerant nature of the SDW phase, the experimental evidences suggest that the value of the ordered magnetic moment at the Fe site in 1111 and 122 pnictides ranges from $0.1\; \mu_{\rm{B}}$ to $1 \; \mu_{\rm{B}}$, i.e., much lower than the value expected in a localized scenario \cite{Dai15}.

However, the phenomenology described above is not universal for all the iron-based materials. In general, it is known that the fermiology of iron-based chalcogenides does not include hole pockets at the $\Gamma$ point, suppressing the nesting effects which are relevant in pnictides. At the same time, several studies have shown that the relative degree of electronic correlations vs.~kinetic energy in iron-based materials is highly variable depending on the actual compound being considered. It is widely accepted that correlations in parent 1111 and 122 pnictides are of moderate intensity -- leading to a ``bad-metallic'' state characterized by high values of electrical resistivity and enhanced effective electron mass -- yet not strong enough to bring the systems to a Mott-insulating state \cite{Si16,Si08,Si09,Qaz09,Abr11,Yin11}. The situation is different for chalcogenides, where the electronic correlations are much more intense relatively to the kinetic energy \cite{Si16,Yin11}. In the exemplary case of the oxychalcogenides  La$_{2}$O$_{3}$Fe$_{2}$(Se,S)$_{2}$, the relative intensity of correlations is so strong that a localized Mott-insulating phase is realized \cite{Zhu10,Gun14}. As a general result, the magnetic phase is qualitatively different in pnictides and in chalcogenides, in terms both of the microscopic magnetic arrangements and of the higher values observed for both $T_{\rm{N}}$ and the ordered magnetic moment \cite{Dag13,Dai15}. In this respect, it is worth pointing out the recent observation by means of NMR of a hedgehog spin-vortex configuration in Ni-doped CaKFe$_{4}$As$_{4}$ \cite{Cui17,Din17}.

The multi-orbital fermiology of iron-based materials leads to exotic physical properties \cite{DeM09,DeM14,DeM15}. When the intensity of electronic correlations get sizeable, the local atomic physics starts to be relevant and the Hund's coupling favours the single electron orbital filling and the decoupling of the charge correlations among the different bands. Accordingly, a significantly different behaviour of the electrons pertaining to different orbitals can be observed, namely an orbital selective behaviour, and eventually an orbitally-selective Mott transition can be detected upon approaching the band half-filling. In particular, for a given stoichiometry, the electrons on $d_{xy}$ bands are characterized by a lower quasiparticle weight -- corresponding to a larger effective mass -- while electrons on $d_{z^{2}}$ or $d_{x^{2}-y^{2}}$ orbitals are characterized by lower effective masses. This behaviour can be associated with a different effective filling of the different bands at a given stoichiometry, approaching the unity for the electrons on the most correlated bands.

\subsection{Effect of chemical substitutions} 
\begin{figure}
	\vspace{15.6cm} \includegraphics{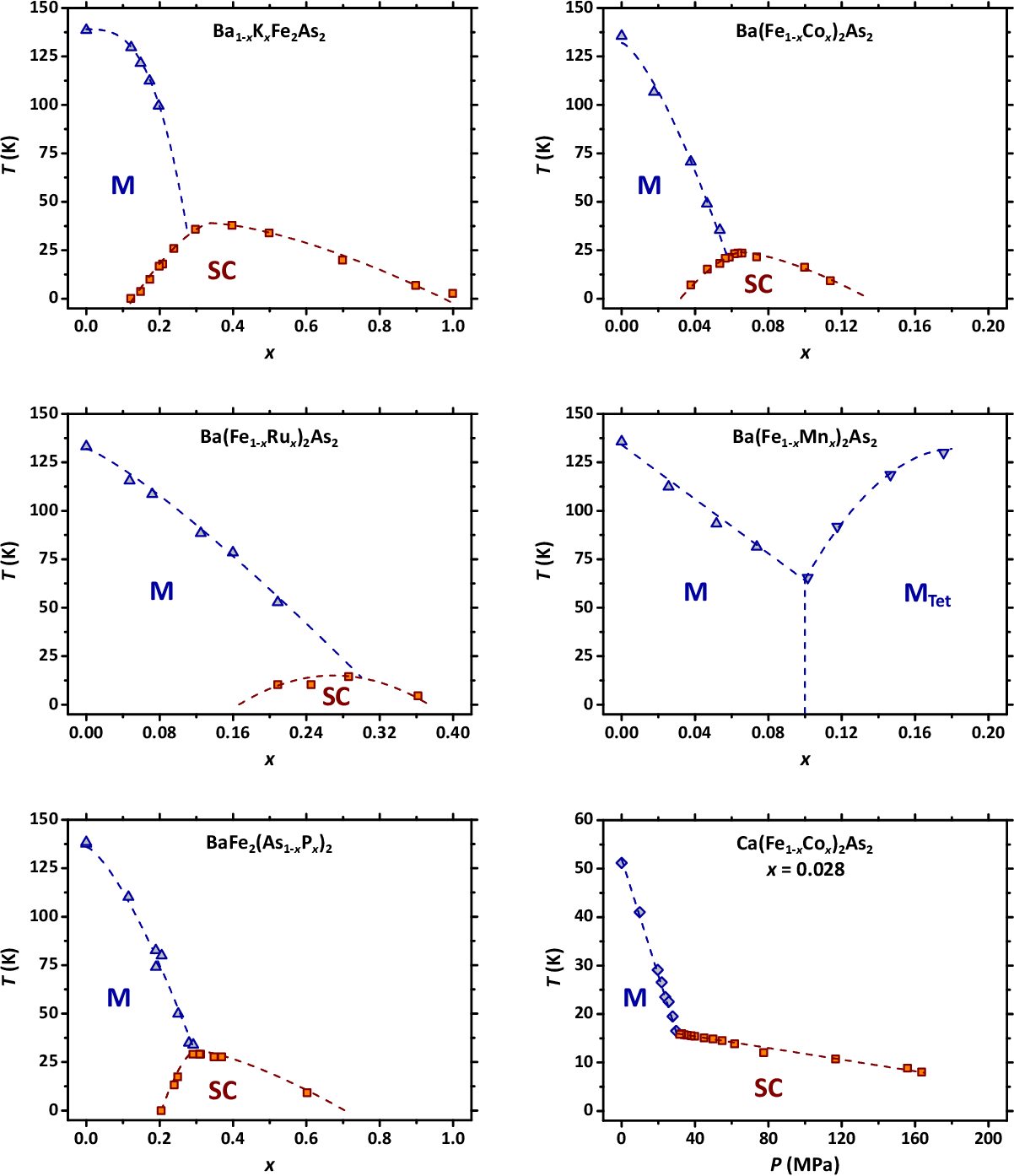} \caption{Electronic phase diagrams for representative materials belonging to the 122 family under the effect of different chemical dilutions and/or external hydrostatic pressure (see the text for details). The labels ``M'' and ``SC'' refer to the magnetic and superconducting phases, respectively. The up-pointing triangles denote the critical temperature for the M phase estimated by means of neutron scattering experiments. This holds also for the down-pointing triangles which, however, correspond to a magnetic phase developing within a tetragonal crystalline environment -- hence the label ``M$_{\rm{Tet}}$''. The diamonds and the squares denote the critical temperature for the M and SC phases, respectively, estimated by means of dc magnetometry. The data for Ba$_{1-x}$K$_{x}$Fe$_{2}$As$_{2}$ are taken from \cite{Avc11}. The data for Ba(Fe$_{1-x}$Co$_{x}$)$_{2}$As$_{2}$ and Ba(Fe$_{1-x}$Ru$_{x}$)$_{2}$As$_{2}$ are taken from \cite{Kim11b}, where the data are partially taken from previous articles in turn. The data for Ba(Fe$_{1-x}$Mn$_{x}$)$_{2}$As$_{2}$ are taken from \cite{Kim10}. The data for BaFe$_{2}$(As$_{1-x}$P$_{x}$)$_{2}$ are taken from \cite{All14} and \cite{Hu15}. The data for Ca(Fe$_{1-x}$Co$_{x}$)$_{2}$As$_{2}$ ($x = 0.028$) are taken from \cite{Gat12}.}\label{FigDilutions}
\end{figure}

The most spectacular physical properties of iron-based materials develop upon perturbing the parent compositions by means of different approaches. This is shown in Fig.~\ref{FigDilutions} for several chemical substitutions and external hydrostatic pressure acting on materials belonging to the 122 family. In particular, the SDW phase developing in the parent compound is progressively suppressed upon increasing $x$ and eventually, close to the suppression of the magnetic state, superconductivity emerges with a characteristic dome-shaped dependence of $T_{\rm{c}}$ on $x$. The fine details of the crossover region between magnetism and superconductivity -- phase segregation, nanoscopic or atomic coexistence -- are often material dependent, and local-probe techniques such as NMR and muon spin spectroscopy are very useful in order to unravel these issues case by case \cite{Lap09,Pra13phd,Car13}. Remarkably, the phenomenology is qualitatively similar regardless of the crystallographic position affected by the dilution -- at variance with the less versatile scenario observed in cuprates.

In general terms, progressive chemical substitutions can have several consequences in condensed matter. The most natural effect is the introduction of quenched disorder, normally with detrimental repercussions on superconductivity. Fig.~\ref{FigDilutions} is exemplary in this sense -- chemical substitutions on the Fe-As layers lead to lower maximum $T_{\rm{c}}$ values if compared to the result of substitutions outside the layers. Also, the examination of Fig.~\ref{FigDilutions} would suggest that the SDW phase is suppressed faster by in-plane substitutions, however this does not look like a common trend for iron-based materials \cite{Pra13}.

Additionally, chemical dilutions normally generate internal strains leading to changes in the structural parameters (chemical pressure) but they may also introduce or remove additional free electrons and/or localized magnetic moments (charge and magnetic doping, respectively). Such a complicated interplay of different effects can be partially disentangled in specific cases -- for example, isovalent substitutions such as As$_{1-x}$P$_{x}$ give the possibility of studying the effect of disorder and chemical pressure alone, and this strategy has been used extensively for iron-based pnictides. In this context, an intriguing result is that the development of the magnetic phase looks correlated with structural parameters such as the distance of the pnictogen element from the Fe layer (the so-called pnictogen height $h_{\rm{Pn}}$), the distance between Fe ions and the Pn--Fe--Pn angle within the distorted FePn$_{4}$ tetrahedra regardless of the actual family of compounds \cite{Sak19}. Naively, the disruption of the SDW phase can be understood as a progressive distortion of the energy bands caused by the structural modifications leading to a departure from the good nesting of the Fermi surface. The more general relevance of the structural parameters for the electronic properties of iron-based materials is made clear by showing that the application of external pressure suppresses the SDW phase and, eventually, induces bulk superconductivity in undoped materials \cite{DeR12,Ali09,Kot09,Kim09}.

The common case of non-isovalent chemical substitutions affects the occupation of the electronic bands as well and both the situations of electron- and hole-doping can be realized, as shown in Fig.~\ref{FigDilutions}. In 1111 materials, the out-of-plane O$_{1-x}$F$_{x}$ dilution has been used frequently to realize electron doping (however, see \cite{Ber19}) even though it is not possible to obtain spatially-homogeneous doping levels for $x > 0.15$. This problem has been circumvented by using the O$_{1-x}$H$_{x}$ dilution, also realizing electron doping and making it possible to explore the electronic phase diagram up to much higher $x$ values. In particular, this strategy made it possible to resolve a bipartite magnetic phase and a second superconducting dome in LaFeAsO$_{1-x}$H$_{x}$ \cite{Hir14}. In general, 122 materials offer a much wider chemical flexibility for the possible substitutions resulting in electron- or hole-doping \cite{Man10}. A common choice is that of substituting transition-metal elements Tm for Fe, the Fe$_{1-x}$Co$_{x}$ substitution having been explored as source of electron doping for 1111 materials as well \cite{Sef08}. Interesting correlations between the structural parameters and the superconducting properties have been reported also for charge-doped materials, pointing towards an optimal $T_{\rm{c}}$ value for specific values of $h_{\rm{Pn}}$ \cite{Miz10}. Also in the case of charge-doped materials, the application of external pressure can mimic the effect of further charge doping \cite{Gat12,Kha11,Pra15}.

A simplified approach to the description of charge doping is the so-called rigid-band approximation. Within this framework, the band structure of parent compounds is assumed to be not altered by the charge doping, which only affects the band occupancy and the position of the Fermi energy $E_{\rm{F}}$ in turn. It is then immediate to realize that a change in $E_{\rm{F}}$ can easily affect the relative sizes of hole and electron pockets and destabilize a condition of good nesting for the Fermi surface, and suppress the SDW phase upon increasing the doping level. In fact, in view of the discussion above concerning isovalent substitutions, more realistic approaches should consider the combined effect of charge doping and structurally-induced tweaks in the band structure -- however, this may well turn out to be too simplistic as well. It is remarkable that a common scaling for the suppression of the SDW state induced by the Fe$_{1-x}$Tm$_{x}$ substitution can be realized by considering the amount of dilution $x$ rather than the effective charge doping \cite{Ni10}. Also, nominally isovalent substitutions such as Fe$_{1-x}$Ru$_{x}$ lead to completely different phase diagrams for LaFe$_{1-x}$Ru$_{x}$AsO and Ba(Fe$_{1-x}$Ru$_{x}$)$_{2}$As$_{2}$, with superconductivity developing only in the latter case \cite{Kim11,Bon12,Ret17}. Whether Fe$_{1-x}$Tm$_{x}$ substitutions actually induce charge doping at all has been highly debated both on the experimental and on the computational sides \cite{Wad10,Bit11,Ber12,Mer12,Lev12,Mer16}. These examples demonstrate the failure of naive approaches to problems such as chemical dilutions and the necessity of considering them in their full complexity, taking into account the role of electronic correlations and Hund's coupling, which become particularly relevant upon hole-doping the 122 compounds. In particular, the decrease in the number of electrons per Fe leads to half-band filling (5e$^{-}$/Fe) where a Hubbard insulating phase is expected, as in the parent compounds of the cuprates. Remarkably, as mentioned above, the electrons in the different bands do not behave in the same way on approaching half filling and an orbital selective behaviour emerges in this regime.

\subsection{Superconducting phase} Small values for the superconducting coherence length and large penetration depths -- typically $\xi \sim 1 - 10$ nm and $\lambda \sim 0.1 - 1 \; \mu$m, respectively -- characterize IBSs as type-II superconductors. Additionally, typical values for the upper critical field $H_{\rm{c}2}$ are as high as $10^{6}$ Oe \cite{Pal12}. In spite of the lower $T_{\rm{c}}$ values, all these properties closely resemble those of cuprate superconductors. However, it is worth mentioning that IBSs are generally characterized by lower values for the anisotropy parameter, determined by the ratio of the in-plane and out-of-plane effective masses, $\gamma_{\rm{s}} \sim 1 - 5$ \cite{Pra11}, to be compared to $\gamma_{\rm{s}}$ values even up to $90$ for cuprates. The low superconducting anisotropy of IBSs has important implications for the engineering of polycrystalline bulk materials capable of sustaining high values of global critical currents \cite{Hos18}.
\begin{figure}
	\vspace{6.1cm} \includegraphics{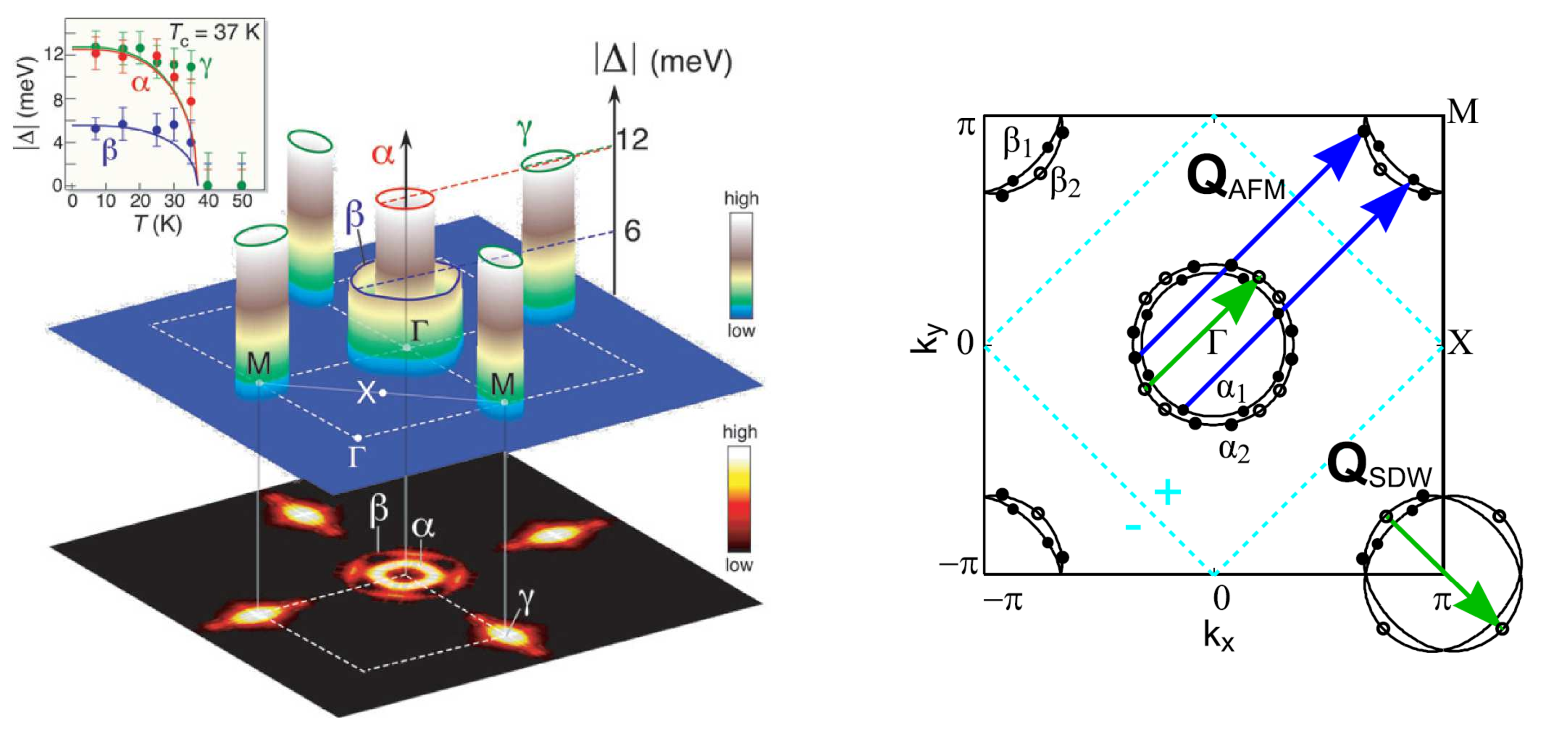} \caption{Left-hand panel: amplitude of the superconducting gaps in the different regions of the Fermi surface of hole-doped Ba$_{1-x}$K$_{x}$Fe$_{2}$As$_{2}$ measured by means of angle-resolved photoemission spectroscopy ($\alpha$, $\beta$: hole-like sheets; $\gamma$; electron-like sheet). The inset shows the evolution of the gaps' amplitudes as a function of temperature, evidencing a saturating behaviour characteristic of $s$-wave symmetry. \{Reprinted figure with permission from \cite{Din08}\}. Right-hand panel: sketchy representation of the relative sign-change of the superconducting order parameter among bands of hole-like or electron-like character in the different regions of the BZ ($s^{\pm}$ symmetry). The vector \textbf{Q}$_{\rm{AFM}}$ represents the inter-band interaction assisted by spin fluctuations and is highly reminiscent of the nesting vector for the Fermi surface within the magnetic phase. \{Reprinted figure with permission from \cite{Kor08} (http://dx.doi.org/10.1103/PhysRevB.78.140509). Copyright (2008) by the American Physical Society.\}.}\label{FigsPlusMinus}
\end{figure}

The most remarkable similarity between IBSs and cuprates is that the superconducting phase emerges in close proximity to the suppression of an antiferromagnetic phase, leading to fundamental questions about the possible interplay between these two states. In fact, several theoretical proposals have been put forward in the last decades addressing the role of antiferromagnetic spin fluctuations as glue for the Cooper pairing in most unconventional superconductors, ranging from cuprates to heavy-fermions \cite{Sca99,Sca12}. This framework has been extended to IBSs soon after their discovery in view of some interesting observations. First of all, the typical values for the electron-phonon coupling are too weak to justify a conventional pairing \cite{Boe08}, although signatures of isotope effects suggest at least a partial role of the lattice in the superconducting mechanism \cite{Kha17}. Most remarkably, inelastic neutron scattering experiments in materials belonging to different families have demonstrated the preservation within the superconducting state of magnetic resonant modes characterized by wavevectors consistent with those leading to the SDW instability \cite{Lum10,Dai15,Kot12,Ino16}. This is normally considered as a strong experimental evidence of a common origin of both the electronic phases which is at the hearth of several theoretical models accounting for high-$T_{\rm{c}}$ superconductivity in IBSs \cite{Maz08,Kur08,Maz09,Hir11,Chu12,Hir16}. In particular, the spin-fluctuation scenario would allow the development of a (possibly anisotropic) $s$-wave gap around the different portions of the Fermi surface with the further assumption of a relative sign-change for the superconducting order parameter among bands of hole-like or electron-like character. This so-called $s^{\pm}$ symmetry is ultimately made possible by the repulsive character of spin fluctuations as well as the disconnected topology of the Fermi surface (see Fig.~\ref{FigsPlusMinus}). This latter property marks a major difference with cuprates, where the spin fluctuations -- acting within a connected Fermi surface instead -- would impose different constraints on the sign-changes of the gap, leading to the so-called $d$-wave symmetry. At the same time, the multi-band $s$-like symmetry makes IBSs closer to high-$T_{\rm{c}}$ multi-band superconductors such as MgB$_{2}$ \cite{Maz10}.

The model outlined above is neither the unique theoretical proposal to describe the mechanism underlying superconductivity in IBSs nor is it universal for all IBSs. In fact, it has been proposed that orbital fluctuations would also lead to high-$T_{\rm{c}}$ superconductivity and, in view of the attractive-like character of the interaction, the arising multi-gap scenario would not require relative sign changes among different portions of the Fermi surface ($s^{++}$ symmetry) \cite{Kon10,Ona15}. On the other hand, the major changes in the fermiology induced by large amounts of charge doping may also lead to a destabilization of the $s$-wave symmetry in favour of a $d$-wave nodal symmetry. This latter aspect has been investigated extensively from both experimental and computational approaches \cite{Fuk09,Zha10,Don10,Ter10,Don10b,Has10,Rei12,Tho11,Mai11}. Several experiments have also shown that changes in the superconducting symmetry can be induced by pressure as well, in agreement with the theoretical proposals suggesting a direct role of structural parameters -- e.g., $h_{\rm{Pn}}$ -- as switches for the superconducting gap symmetry \cite{Kur09,Taf13,Gug15}.

\subsection{Electronic nematicity}

A peculiar feature of the microscopic magnetic stripy arrangement sketched in Fig.~\ref{FigAFM} is that the two crystallographic in-plane directions $[1 0 0]_{\rm{O}}$ and $[0 1 0]_{\rm{O}}$ are inequivalent -- indeed, the magnetic coupling is ferromagnetic along one direction and antiferromagnetic along the other. The high-temperature phase is tetragonal and there is no way to associate \textit{a priori} each direction with a specific magnetic coupling. Accordingly, two symmetries are broken upon cooling the system from the high-temperature tetragonal paramagnetic phase to the low-temperature magnetic state, i.e., the Ising symmetry $Z_{2}$ -- associated to the choice of the axis sustaining a ferromagnetic/antiferromagnetic coupling -- and the spin-rotational symmetry $O(3)$ leading to the magnetically-ordered state \cite{Fer12}. The spontaneous breaking of the two symmetries is almost simultaneous in undoped BaFe$_{2}$As$_{2}$, however these develop as distinct phase transitions in Ba(Fe$_{1-x}$Co$_{x}$)$_{2}$As$_{2}$ upon increasing $x$ \cite{Chu09}. Experimentally, this is reflected in the observation of a tetragonal-to-orthorhombic structural phase transition at $T_{\rm{s}}$ breaking the Ising symmetry $Z_{2}$ without necessarily breaking $O(3)$ which, eventually, is broken at $T_{\rm{N}}$. This observation is not specific to Co-doped BaFe$_{2}$As$_{2}$ but it is general for most iron-based materials \cite{Mar16}. In fact, already in undoped LaFeAsO and in undoped NaFeAs it is observed that $T_{\rm{s}} > T_{\rm{N}}$ \cite{Lue09,Hes16,Zho16,Bae18,Toy18}. A limit situation is observed in FeSe, where the structural transition develops in the absence of the magnetic one \cite{Boh16,Boh18}. As a result, an orthorhombic paramagnetic state can be observed over a wide temperature range in most-iron based materials. This phase is referred to as nematic after a state of liquid crystals where the rotational -- but not the translational -- symmetry is broken \cite{Cha95,Fra10}. Correspondingly, the structural transition is often referred to as nematic transition.

As mentioned above, the structural transition has minor quantitative effects on the crystallographic properties. However, it is remarkable that characteristic electronic properties exhibit a marked in-plane anisotropy in the nematic state \cite{Chu10,Fis11} which is hard to be justified in terms of a structural origin. Accordingly, several theoretical proposals have been put forward accounting for nematicity as the result of an electronic instability. The two main proposed scenarios involve spin and orbital degrees of freedom as driving force for nematicity and, at the time of writing, it is not settled yet which picture is the correct one \cite{Fer14}. This is reminiscent of what is discussed above concerning the models accounting for the development of superconductivity, and in fact most of the interest in unravelling the origin of the nematic state is its close relationship to that of high-temperature superconductivity \cite{Fer14}.

\section{Nuclear magnetic and quadrupolar resonance}

NMR and NQR have played a key role in the understanding of the normal state excitations and of the superconducting state. In this section we shall introduce the very basic quantities which can be measured with these techniques in order to allow a deeper understanding of the subsequent sections dedicated to the findings of NMR and NQR in iron-based superconductors.

\subsection{Basic principles} NMR and NQR are spectroscopies relying on the use of nuclei as local probes of their nanoscopic environment \cite{Abr61,Abr70,Fuk81,Ern87,Sli90,Fri98,Rig98,Car07,Wal08}. In very general terms, all those nuclei characterized by a spin angular momentum $\hbar I$ with $I \geq \frac{1}{2}$ possess a non-vanishing magnetic moment $\bm{\mu}_{I} = \gamma \hbar \bm{I}$, with $\gamma$ the gyromagnetic ratio characteristic for every nucleus (see Tab.~\ref{TabNMRNuclei}). Accordingly, nuclei with non-vanishing magnetic moment can interact with magnetic fields and can be used as quantum sensors of the local magnetic, structural and electronic properties of matter. Additionally, all those nuclei characterized by a spin $I \geq 1$ also possess a non-vanishing electric quadrupole moment $Q$ -- see Tab.~\ref{TabNMRNuclei} -- which can interact with the EFG (electric field gradient) generated by the local distribution of electric charge.

\begin{table}[b!]
\caption{Relevant physical quantities of selected nuclei of interest in the study of IBSs. $\gamma$ is the gyromagnetic ratio while $\mu_{I}$ is the magnetic moment associated with the spin I (in units of the nuclear magneton $\mu_{N}$). $Q$ quantifies the electric quadrupolar moment.}\label{TabNMRNuclei}
	\begin{center}
		\begin{tabular}{c|ccccc}%
			{Nucleus} & Natural abundance& $\gamma/2\pi$ & $I$ & $\mu_{I}$ & $Q$ \\
			& (\%) & (MHz T$^{-1}$) & & ($\mu_{N}$ units) & (barn)\\%
			\hline
			{}$^{19}$F & $100$ & $40.0616$ & $1/2$ & $2.6269$ & $0$\\
			{}$^{31}$P & $100$ & $17.2352$ & $1/2$ & $1.1316$ & $0$\\
			{}$^{75}$As & $100$ & $7.2901$ & $3/2$ & $1.4395$ & $0.314$\\
			{}$^{139}$La & $99.91$ & $6.0141$ & $7/2$ & $2.7830$ & $0.2$\\
		\end{tabular}
	\end{center}
\end{table}
The NMR/NQR signal describes the time evolution of the nuclear magnetization as influenced by different interactions, ranging from the dipolar magnetic coupling among different nuclear moments ($\mathcal{H}_{\rm{NN}}$) -- both like and unlike -- to the interaction of the nuclear quadrupole moment with the EFG associated with the crystal electric field ($\mathcal{H}_{\rm{Q}}$) and to the hyperfine interaction of nuclear and electronic magnetic moments ($\mathcal{H}_{\rm{NE}}$). Accordingly, the resulting nuclear spin Hamiltonian $\mathcal{H}_{\rm{N}}$ is generally written as
\begin{equation}\label{EqTotalNuclearHam}
	\mathcal{H}_{\rm{N}} = \mathcal{H}_{\rm{Z}} + \mathcal{H}_{\rm{Q}} + \mathcal{H}_{\rm{NE}} + \mathcal{H}_{\rm{NN}}
\end{equation}
where $\mathcal{H}_{\rm{Z}}$ is the Zeeman term of interaction with external or internal magnetic fields. The diagonalization of $\mathcal{H}_{\rm{N}}$ leads to a set of eigenvalues for the nuclear system and a typical NMR experiment is designed in such a way as to induce transitions among these energy levels resonantly with a radio-frequency magnetic field, subject to the relevant magnetic dipole selection rules.

In NMR, $\mathcal{H}_{\rm{Z}}$ is the dominant term in Eq.~\ref{EqTotalNuclearHam} and the quantization axis -- conventionally aligned with $z$ -- is set by the direction of the magnetic field $\bm{H}_{0}$. In the idealized limit $\mathcal{H}_{\rm{N}} = \mathcal{H}_{\rm{Z}}$, one has
\begin{equation}\label{EqZeemanHam}
	\mathcal{H}_{\rm{Z}} = - \bm{\mu}_{I} \cdot \bm{H}_{0} = - \hbar \gamma \bm{I} \cdot \bm{H}_{0}
\end{equation}
leading to energy levels defined by the quantum number $m_{I}$, i.e., the projection of $\bm{I}$ along $\bm{H}_{0}$. The separation between adjacent levels turns out to be $\Delta E = \hbar \gamma H_{0} = \hbar \omega_{0}$, with $\omega_{0}$ the Larmor frequency. Thus, if $\mathcal{H}_{\rm{Z}}$ is the only term of the Hamiltonian, a single NMR line at $\nu_{0}= \omega_{0}/2 \pi$ is detected.

On the other hand, $\mathcal{H}_{\rm{Q}}$ is the dominant term in NQR and the principal axes of the EFG tensor define the quantization axis. An ideal physical system governed by $\mathcal{H}_{\rm{N}} = \mathcal{H}_{\rm{Q}}$ would then lead to the Hamiltonian
\begin{equation}\label{EqQuadrupolarHam}
	\mathcal{H}_{\rm{Q}} = \frac{eQV_{zz}}{4I(2I-1)} \left[\left(3I_{z}^{2}-I^{2}\right) + \eta \left(I_{x}^{2}-I_{y}^{2}\right)\right] \qquad \left(\eta = \frac{V_{xx}-V_{yy}}{V_{zz}}\right)
\end{equation}
where $V_{xx}$, $V_{yy}$ and $V_{zz}$ are the principal components of the EFG tensor verifying the Laplace equation $\sum_{\alpha} V_{\alpha\alpha} = 0$ ($\alpha = x,y,z$). Additional relations between these components reflect the crystallographic symmetries of the system under investigation. For the specific case of iron-based materials in the tetragonal phase, the principal axes of the EFG tensor at the $^{75}$As nuclei coincide with the crystallographic axes and the relation $|V_{zz}| > |V_{xx}|$ holds as well as the equality $V_{xx} = V_{yy}$. This latter condition is broken by the small structural transition to the orthorhombic phase. These structural changes can be described by the asymmetry parameter $\eta$ (see Eq.~\ref{EqQuadrupolarHam}), which is zero when the EFG tensor is characterized by a cylindrical symmetry and non-zero when such symmetry is lowered. For $I = 3/2$ the above Hamiltonian leads to a single NQR resonance line at
\begin{equation}\label{nqrspe}
	\nu_{\rm{Q}}= \frac{eQV_{zz}}{2h} \left(1+ \frac{\eta^{2}}{3}\right)^{1/2} \,\,\, .
\end{equation}
Thus, NQR is an important tool to monitor locally the development of abrupt structural phase transitions \cite{Rig84} through a splitting or a shift of the line. At the same time, sudden modifications of the EFG may also stem from the development of ordered states for the distribution of electrical charges, such as in charge-density wave phases \cite{Wu11}.

\subsection{Shift and broadening of the resonance line} Variations in the external thermodynamical parameters lead to valuable information on the crystalline environment of the nuclei. For instance, changes in temperature and pressure affect the crystalline structure and the induced progressive shrinkages/expansions are reflected directly by changes in the EFG, readily detected by quadrupolar nuclei. At the same time, these parameters may also influence the local static uniform spin susceptibility $\chi(\bm{q} = 0, \omega = 0)$ [or, in short, $\chi(0,0)$] which affects the nuclei in turn via $\mathcal{H}_{\rm{NE}}$, causing a change in the local field at the nuclei with respect to $\bm{H}_{0}$. Hence, the resonance frequency $\omega_{\rm{r}}$ is shifted with respect to the Larmor frequency $\omega_{0} = \gamma H_{0}$. The paramagnetic shift $\Delta K$ (often called Knight shift in metals) allows one to measure the local static uniform spin susceptibility. In fact, by taking $\mathcal{H}_{\rm{NE}}= \gamma\hbar \bm{I} \bm{\mathcal{A}} \bm{S}$, with $\bm{\mathcal{A}}$ the (tensorial) hyperfine coupling of the nuclear spin $I$ with the electronic spin $S$, one can write 
\begin{equation}\label{shift}
	\Delta K = \frac{\omega_{\rm{r}} - \omega_{0}}{\omega_{0}}= \frac{\mathcal{A} \langle S\rangle}{H_{0}}= \mathcal{A} \chi(0,0)
\end{equation}
in the linear response regime. Long-range ordered magnetic phases may lead to significant shifts or even splittings of the resonance line, the details depending, for example, on the orientation of the internal magnetic field with respect to the quantization axis and on the degree of commensurability of the ordered phase with the underlying lattice symmetry.

Every possible transition between a couple of energy levels is associated with a finite lifetime $\tau_{\rm{coh}}$. Under ideal conditions, this quantity is measurable in an experiment as an intrinsic broadening of the resonance condition around the resonant frequency by an amount $\Delta \nu \sim \tau_{\rm{coh}}^{-1}$. When performing an experiment on a real sample, one deals with a macroscopic amount of nuclei and the quantum mechanical conditions of all the nuclei must be rigorously identical and homogeneous across the sample if one needs to detect the intrinsic broadening. This is the reason why the intrinsic term $\Delta \nu$ is conventionally referred to as homogenous broadening. A prototypical, temperature-independent source of homogeneous broadening for NMR resonance lines is given by the dipolar interaction between like nuclei -- i.e., between the same isotopes. However, in real samples, the microscopic interactions often lead to inhomogeneous conditions mostly associated with the disorder and/or the anisotropic nature of the couplings. The situation is particularly severe in solids and, as a result, the broadening normally detected for resonance lines in solids is often much larger than the limit set by $\tau_{\rm{coh}}^{-1}$. This condition is referred to as inhomogeneous broadening. Typically, the interaction with unlike nuclei and paramagnetic centres cause an inhomogeneous broadening. In type-II superconductors, the modulation of local magnetic fields induced by vortices in the mixed Shubnikov state is an example of temperature-dependent source of inhomogeneous broadening (see Sect.~\ref{Sectflux}), from which interesting information can be extracted \cite{Car92}. The temperature dependence of the linewidth of the NMR spectra is associated both with the temperature dependence of the magnetic field penetration depth as well as from the motional narrowing of the spectra induced by the flux lines motions. Extrinsic sources of broadening arise also from the random orientation of the crystallites in powder samples. For each crystallite orientation a different NMR resonance frequency can be expected.

\subsection{Spin-lattice and spin-spin relaxation times}\label{SectT1T2}
Considering a macroscopic nuclear system described by Eq.~\ref{EqTotalNuclearHam}, the populations of the energy levels at the thermodynamical equilibrium are described by Boltzmann statistics. The resulting configuration is characterized by an equilibrium value of the nuclear magnetization $\bm{M}_{I}^{\rm{eq}}$ that can be calculated from a Curie-like treatment of the paramagnetic susceptibility. The statistical occupancy of the energy levels makes it possible to define a spin temperature $T_{\rm{spin}}$ that, under thermodynamical equilibrium conditions, is equal to the temperature $T_{\rm{L}}$ of the surrounding medium -- conventionally referred to as the ``lattice''. Sudden changes in the conditions of the external parameters (e.g., reversals of the magnetic field) or suited pulsed protocols can bring the system out of equilibrium and $T_{\rm{spin}}$ can take values from $-\infty$ to $+\infty$. The characteristic time by which the population on the energy levels goes back to equilibrium is $T_{1}$, the spin-lattice relaxation time. $T_{1}$ is directly related to the transition probability among the energy levels driven by the time-dependent part of $\mathcal{H}_{\rm{N}}$. Since in NMR, for $\bm{H}_{0} // z$, the population difference between the energy levels is proportional to $(M_{I})_{z}$, $T_{1}$ is determined by the recovery of $(M_{I})_{z}$ back to its thermal equilibrium value. If the time-dependent part of $\mathcal{H}_{\rm{N}}$ is associated with fluctuations of the local magnetic field $\bm{h}$, determined for example by electron spin fluctuations or by the flux lines dynamics, then
\begin{equation}\label{T1h}
	\frac{1}{T_{1}} = \frac{\gamma^{2}}{2} \int_{-\infty}^{+\infty} \langle h_{+}(t)h_{-}(0) \rangle e^{-i\omega_{\rm{r}} t} dt \,\,\, ,
\end{equation}
where $h_{\pm}(t) = h_{x}(t) \pm i h_{y}(t)$. The Fourier transform at the resonance frequency is involved owing to the conservation of energy, while the magnetic dipole selection rules involve only the components of the fluctuating field perpendicular to the quantization axis. It is often possible to write $\langle h_{+}(t)h_{-}(0) \rangle = \langle \Delta h_{\perp}^{2} \rangle \exp(-t/\tau_{\rm{c}})$ and accordingly, based on Eq.~\ref{T1h},
\begin{equation}\label{BPP}
	\frac{1}{T_{1}}= \gamma^{2} \langle \Delta h_{\perp}^{2} \rangle \frac{\tau_{\rm{c}}}{1 + \omega_{\rm{r}}^{2}\tau_{\rm{c}}^{2}}
\end{equation}
leading to a peak in $1/T_{1}$ when the frequency of the fluctuations $\omega_{\rm{c}} = 1/\tau_{\rm{c}}$ matches the resonance frequency $\omega_{\rm{r}}$. If the fluctuations are driven by electron spins localized at lattice sites labelled by the index $j$, one can write
\begin{equation}
	 \bm{h}(t) = \sum_{j} \bm{\mathcal{A}}_{j}\bm{S}_{j}(t) = \frac{1}{\sqrt{N}} \sum_{\bm{q}} \sum_{j} \mathcal{A}_{j} \bm{S}_{\bm{q}}(t) e^{i \bm{q}\cdot\bm{r}_{j}}
\end{equation}
which, after substitution in Eq.~\ref{T1h}, leads to
\begin{equation}\label{T1Sqw}
	\frac{1}{T_{1}}= \frac{\gamma^{2}}{2} \frac{1}{N}\sum_{\bm{q}} \left[|A_{\bm{q}}|^{2} S_{\alpha\alpha}(\bm{q},\omega_{\rm{r}})\right]_{\perp} \,\,\, .
\end{equation}
Here, $|A_{\bm{q}}|^{2}$ is the form factor giving the hyperfine coupling of the nuclei with the spin excitations at the wavevector $\bm{q}$ and $S_{\alpha\alpha}(\bm{q},\omega_{\rm{r}})$ is the component of the dynamical structure factor at the resonance frequency. Accordingly, $1/T_{1}$ is determined by the $\bm{q}$-integrated amplitude of the low-frequency fluctuations, properly filtered by the form factor.

The local fluctuations do not affect only the longitudinal component of the nuclear magnetization but also the transverse component, whose time evolution is determined by the characteristic spin-spin relaxation time $T_{2}$. While $1/T_{1}$ can be defined unambiguously, the estimate of $1/T_{2}$ depends very much on the radio-frequency pulse sequence being used. $T_{2}$ is often defined as the characteristic decay time of the echo after the $\pi/2-\tau -\pi$ Hahn sequence, where $\pi/2$ and $\pi$ are the nuclear spin flip angles. This sequence is very effective in generating an echo as far as the magnetic field is static over a time of the order of $\tau$, otherwise the echo amplitude is reduced. For example, for a correlation function of the fluctuations decaying exponentially over a time $\tau_{\rm{c}}$, the echo amplitude decays according to the following expression ($\tau$ being the separation between the $\pi/2$ and $\pi$ pulses) \cite{Tak86}
\begin{equation}\label{Echo}
	E(2\tau) = E(0) \exp\left\{-\langle \Delta\omega^{2} \rangle \tau_{\rm{c}} \, \left[2\tau - \tau_{\rm{c}} (1 - e^{-\tau/\tau_{\rm{c}}})(3 - e^{-\tau/\tau_{\rm{c}}})\right]\right\}\,\, .
\end{equation} 
The presence of very low-frequency fluctuations can be evidenced by recording echo combs using the CPMG (Carr-Purcell-Meiboom-Gill) pulse sequence with different separations among the radio-frequency pulses \cite{Sli90}.
\begin{figure}
	\vspace{5.2cm} \includegraphics{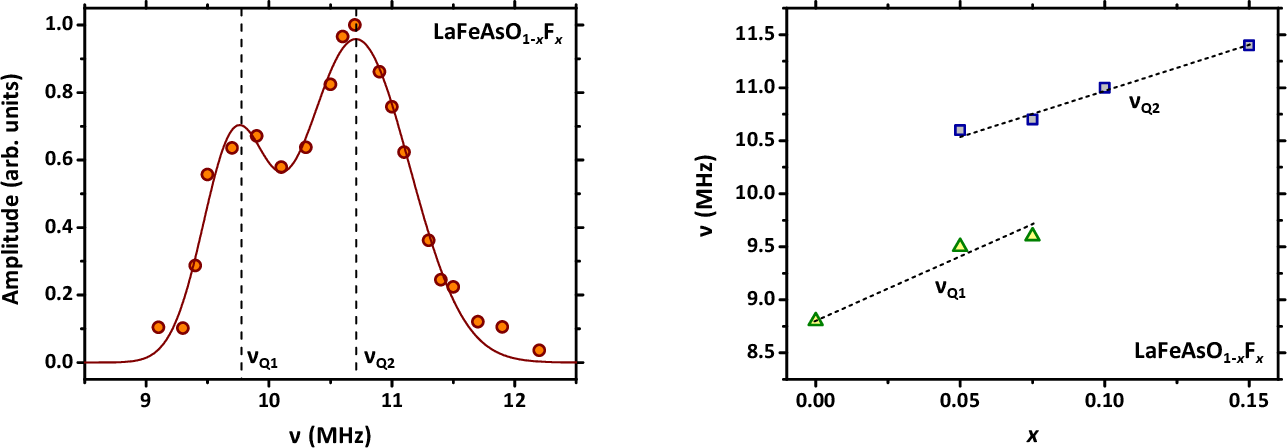} \caption{Left-hand panel: representative $^{75}$As NQR spectrum in an underdoped LaFeAsO$_{1-x}$F$_{x}$ compound. For a nuclear spin $I = 3/2$ a single line is expected for a uniform charge distribution. The presence of two peaks has been associated with the presence of charge-poor and charge-rich regions  segregated at the nanoscale \cite{Lan10}. Right-hand panel: doping dependence of the NQR resonance frequencies in LaFeAsO$_{1-x}$F$_{x}$. Upon increasing the F content, for $0.03\leq x\leq 0.1$, two peaks are observed (green and blue symbols show the corresponding NQR frequencies). In this doping range the NQR frequencies vary only slightly and, upon increasing $x$, the amplitude of the low-frequency peak decreases while the one of the high frequency peak increases \cite{Lan16}. This means that the ratio between the volume of the charge-rich and charge-poor regions gets progressively enhanced by fluorine doping.}\label{FigNQR2peaks}
\end{figure}

\section{A local view on the normal state of iron-based superconductors}\label{Sectnorm}

\begin{figure}
	\vspace{6.0cm} \includegraphics{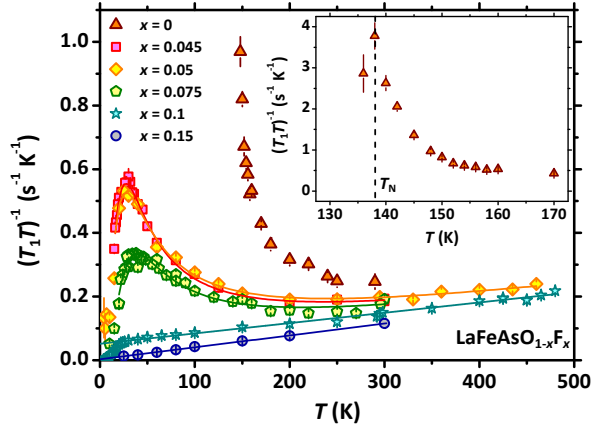} \caption{Temperature dependence of $^{75}$As $1/T_{1}T$ in LaFeAsO$_{1-x}$F$_{x}$ for several compositions at $H_{0} = 70$ kOe. The solid lines are the best fits for a $1/T_{1}$ given by the sum of a power-law describing the metallic behaviour and by Eq.~\ref{BPP} describing low-frequency fluctuations, with a thermally activated correlation time characterized by a distribution of energy barriers \cite{Ham13}. The inset enlarges the divergence of the spin-lattice relaxation rate at the magnetic ordering temperature in LaFeAsO.}\label{T1Franz}
\end{figure}
The different quantities measured in NMR-NQR experiments, i.e., the spectra and the relaxation rates, can be used to investigate the normal state excitations of iron-based materials. In fact, the study of the normal state excitations has always been central for the understanding of the mechanism driving superconductivity. Akin to the cuprates \cite{Kei15}, also the phase diagram of IBSs is characterized by several crossovers and phase transitions evidencing the presence of competing energy scales which generates a rather rich phenomenology, characteristic of strongly correlated electron systems such as heavy fermions \cite{Wir16,Fer19}. Within that complex phenomenology one has to sort out which is the glue for the Cooper pairs and how it is affected by different parameters like the stoichiometry, the pressure and the magnetic field. The competing energy scales characterizing these materials may lead to an electronic phase separation \cite{Dag05}, as evidenced by the $^{75}$As NQR spectra of electron-doped LaFeAsO$_{1-x}$F$_{x}$ and shown in Fig.~\ref{FigNQR2peaks} \cite{Lan10,Lan16}. It has been observed that by introducing electrons through fluorine doping the NQR spectrum not only broadens, as expected in the presence of disorder, but is also composed of two peaks, suggesting the coexistence of two different electronic configurations. The low- (high-)frequency peak is associated with charge-poor (rich) regions. Remarkably, the $1/T_{1}$ measured on both peaks is the same \cite{Lan10}, suggesting that these two regions are segregated at the nanoscale so that nuclear spin diffusion establishes a common spin temperature between $^{75}$As nuclei in both the regions \cite{Blo49}. This inhomogeneous charge distribution develops over a fluorine doping range where the coexistence of magnetism and superconductivity takes place at low temperature. It can be enhanced by the quenched disorder associated with a non-uniform distribution of the fluorine dopant, which can effectively pin the charges in these correlated electron systems. Remarkably, as we shall see in Sect.~\ref{Chargesect}, a non uniform charge distribution -- i.e., a charge order -- is present also in stoichiometric compounds where the quenched disorder is rather low, as demonstrated by the high residual resistivity ratio.

In the most studied electron-doped IBSs -- such as those belonging to 1111 and 122 families -- it has become clear since the early days that, in spite of the presence of weaker electronic correlations with respect to the cuprates, a Fermi liquid description is not appropriate. In fact, clear deviations are observed from the standard Korringa behaviour observed in metals \cite{Sli90}, since both $1/T_{1}T$ and the paramagnetic shift $K_{s}$ are not constant \cite{Ham13,Oh11,Nak08,Ahi08}. Similar results were reported recently for ThFeAsN, a compound isostructural to 1111 materials whose parent composition is superconducting \cite{Shi17,Bar18}. From the analysis of $1/T_{1}$ anisotropy -- for $\bm{H}_{0} //$ or $\bm{H}_{0} \perp c$ -- it was found that spin correlations are peaked at the collinear stripy wavevector $\left(\frac{\pi}{a},0\right)$ or $\left(0,\frac{\pi}{a}\right)$ \cite{Kit10}.

\begin{figure}
	\vspace{6.5cm} \includegraphics{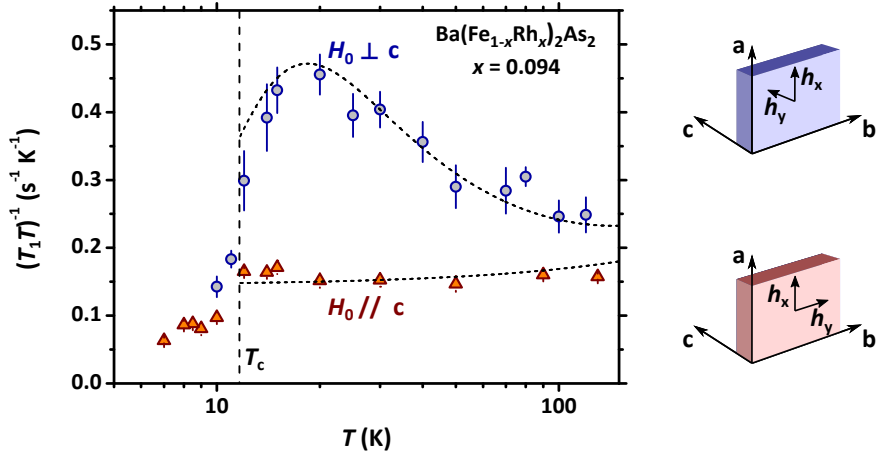} \caption{Temperature dependence of $^{75}$As NMR $1/T_{1}T$ in Rh-doped BaFe$_{2}$As$_{2}$, for a magnetic field $H_{0} = 70$ kOe both along the $c$ axis and in the $ab$ plane (the dashed lines are guides to the eye). For $\bm{H}_{0} // c$, $1/T_{1}T$ shows a nearly-constant behaviour in the normal state, characteristic of weakly correlated metals. For $\bm{H}_{0} \perp c$, $1/T_{1}T$ evidences an additional contribution giving rise to a peak when the fluctuations frequency approach the Larmor frequency. On the right the components of the fluctuating magnetic field involved in the relaxation process are sketched. The vertical dashed line marks the superconducting transition temperature.}\label{FigT1Ba122}
\end{figure}

In the pnictides, where the most used probe is $^{75}$As, one observes a peak in $1/T_{1}T$ at low temperature in the underdoped regime of 1111 compounds (see Fig.~\ref{T1Franz}), as observed in systems approaching a spin freezing \cite{Dio13,Dio15}. Also in the heavily-overdoped regime of 122 compounds a peak is observed, but only for $\bm{H}_{0} \perp c$ (see Fig.~\ref{FigT1Ba122}). Close to optimal doping $1/T_{1}T$ is nearly constant, while upon increasing the electronic doping it is found to decrease upon cooling, a behaviour tentatively ascribed to the onset of a pseudo-gap, namely to a decrease in the density of states at the Fermi level. This latter scenario has been put forward also for compounds such as (CaFe$_{1-x}$Pt$_{x}$As)$_{10}$Pt$_{3}$As$_{8}$ \cite{Sur15}. Remarkably the peak in $1/T_{1}T$ at low electron doping levels shows also a clear magnetic field dependence, its amplitude decreasing upon increasing the magnetic field \cite{Ham13}. This behaviour is typically observed in the presence of a slowing down of the fluctuations approaching the hundreds or tens of MHz, suggesting the presence of unusual low-frequency excitations possibly involving nematic fluctuations (see Sect.~\ref{Sectnem}).

Also the temperature dependence of the paramagnetic shift is not Pauli-like as expected for a Fermi liquid. In the electron-doped 122 compounds, for example, an activated increase of the paramagnetic shift with temperature is observed \cite{Oh11}. Such a behaviour could again be an indication of a pseudo-gap, as suggested by $1/T_{1}$ measurements. Another possibility is that the behaviour originates from the singularities in the density of states which, when their width is of the order of $k_{\rm{B}}T$, cause a temperature dependence of the spin susceptibility. By considering the density of states around the Fermi level derived from DFT (density functional theory) calculations for Rh-doped BaFe$_{2}$As$_{2}$ Bossoni \textit{et al.} have calculated the temperature dependence of the paramagnetic shift \cite{Bos12}. A good agreement with the experimental data can be found only by leaving the Fermi level as a free parameter, which would imply a marked dependence of the paramagnetic shift magnitude on the electron doping level, at variance with the experimental findings. The need to go beyond DFT analysis is a first indication for the relevance of electronic correlations in these materials, even far away from half-band filling.

\subsection{Magnetic and nematic phases}\label{Sectnem}

\begin{figure}
	\vspace{6.5cm} \includegraphics{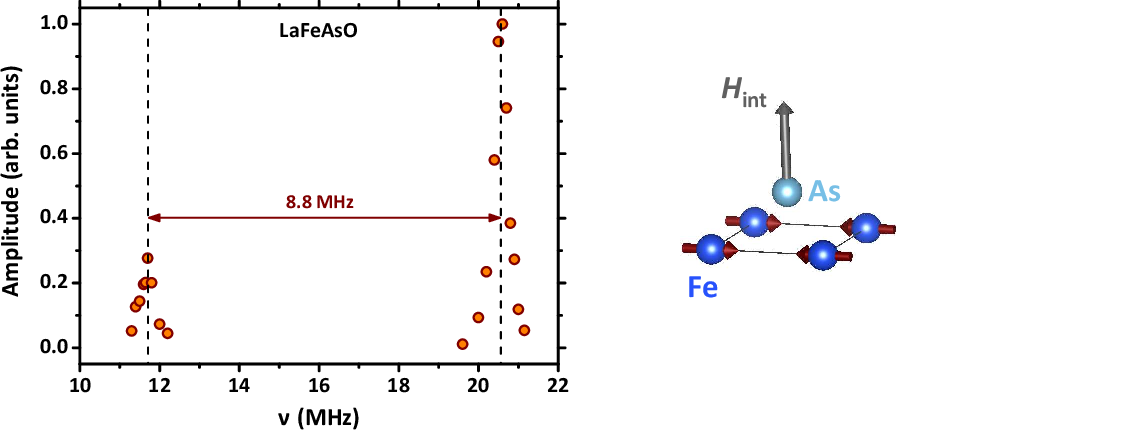} \caption{$^{75}$As zero-field NMR spectrum in LaFeAsO at $T = 4.2$ K -- i.e., in the magnetically ordered phase. The low-frequency peak is associated with the $+1/2\leftrightarrow -1/2$ transition while the high-frequency peak with the $+3/2 \leftrightarrow +1/2$ transition. The horizontal line evidences the quadrupolar splitting $\nu_{\rm{Q}}$. The intensity of the peaks is affected by instrumental factors. On the right, the orientation of the internal hyperfine field at $^{75}$As nucleus is shown.}\label{ZFLaFeAsO}
\end{figure}

If one considers the magnetic parent compounds of IBS, where a SDW order sets in thanks to the Fermi surface nesting, one finds a behaviour of $1/T_{1}$ typical of magnetically ordered compounds with a critical divergence of the spin-lattice relaxation rate at $T_{\rm{N}}$ (see Fig.~\ref{T1Franz}) \cite{Ish09}. As expected, the internal field generated by the electrons at the nuclei shifts the resonance frequencies in the NMR spectra, also in zero-field (see Fig.~\ref{ZFLaFeAsO}). However, a deeper view on the behaviour of the NMR spectra and of $1/T_{1}$ actually evidences two of the most interesting physical aspects of IBS: the presence of a nematic order and hints for an orbital selective behaviour. The nematic order induces a breaking of the tetragonal symmetry and accordingly a change in the EFG at $^{75}$As nuclei, detected by the NMR spectra \cite{Fu12}. While these changes typically occur very close to the magnetic ordering temperature, where the magnetic correlations are stronger, clear evidences for an in-plane anisotropy of the EFG well above $T_{\rm{N}}$ have been presented by Iye \textit{et al.} \cite{Iye15} based on the analysis of the $^{75}$As NMR linewidth in P-doped BaFe$_{2}$As$_{2}$. The possibility to have an in-plane anisotropy extending well above $T_{\rm{N}}$ has been also suggested by Martinelli \textit{et al.}, based on a detailed analysis of neutron and XRD results in 1111 compounds \cite{Mar11}. These effects could certainly be affected by disorder which would tend to pin either orbital or purely magnetic nematic fluctuations.

A significant interest was drawn by the observation of a nematic order in FeSe \cite{McQ09,Boh15,Bae15,Bae16,Wie17} where long range magnetic order is absent, a situation analogous to the one recently observed in RbFe$_{2}$As$_{2}$ \cite{Ish18}. The absence of a magnetic long range order does not rule out the magnetic origin of the nematic order. In fact, as in the $J_{1}-J_{2}$ model on a square lattice \cite{Cha90}, the nematic order can be driven by sufficiently large spin correlations which can induce a breaking of the $C_{4}$ symmetry even without the onset of any long range magnetic order \cite{Car11b}.
\begin{figure}
	\vspace{6.5cm} \includegraphics{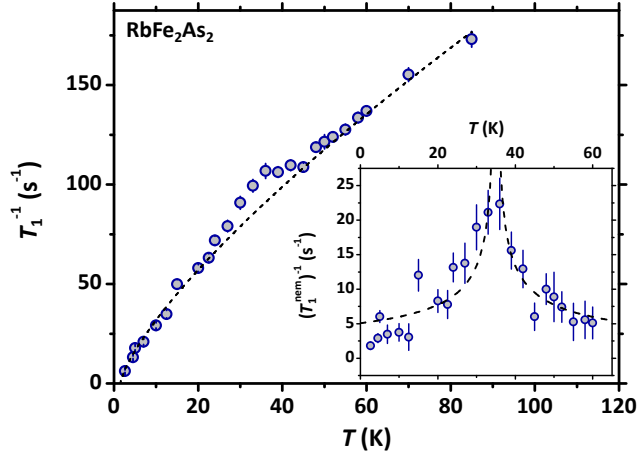} \caption{Temperature dependence of the NQR $^{75}$As $1/T_{1}$ in RbFe$_{2}$As$_{2}$. The dashed line shows the average power-law behaviour on top of which a bump emerges around 35 K. The inset shows the $1/T_{1}$ around that temperature after subtracting the power-law behaviour. The evidenced divergence has been associated tentatively with a nematic order \cite{Mor19}.}\label{XYnemfig}
\end{figure}

The breaking of the $C_{4}$ symmetry is expected to lead to a divergence of the so-called nematic susceptibility which can be probed by perturbing the material with an external uniaxial strain $\epsilon$ \cite{Boh16}. For example, one can measure how the strain modifies the in-plane ($ab$ plane) anisotropy of the electrical resistivity $\Delta\rho_{ab}$ and accordingly probe the effect of the nematic order on the electronic properties. More interestingly, one can probe the effect of a uniaxial strain both on $T_{1}$ and on the EFG asymmetry parameter $\eta$. While the former provides information on the nematic spin correlations the latter yields information on the anisotropy in the electronic charge distribution. In fact, by carefully measuring the effect of a uniaxial strain on $\eta$ in BaFe$_{2}$As$_{2}$, Kissikov \textit{et al.} \cite{Kis16,Kis17} found a divergence of $d\eta/d\epsilon$ which nicely correlates with the divergence of $\chi_{\rm{nem}}= d\Delta\rho_{ab}/d\epsilon$. In the same compound the same group carried out a careful study of the enhancement of $1/T_{1}$ for particular geometries of the magnetic field and strain orientation and derived a spin nematic susceptibility which shows a trend similar to the one derived from elastoresistivity measurements \cite{Kis17b}.
\begin{figure}
	\vspace{6.5cm} \includegraphics{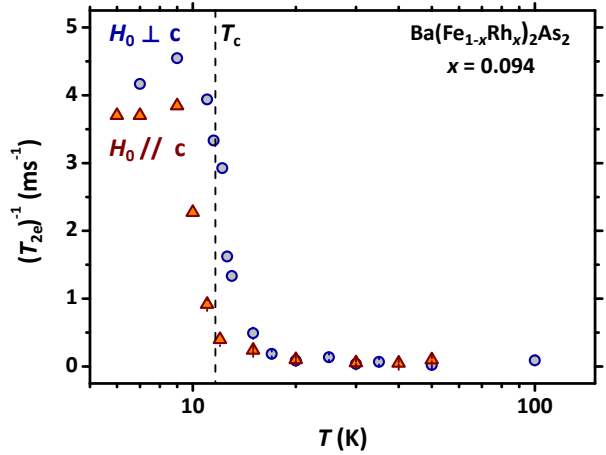} \caption{Temperature dependence of $^{75}$As $1/T_{2\rm{e}}$ in Ba(Fe$_{1-x}$Rh$_{x}$)$_{2}$As$_{2}$ for $x = 0.094$ at $H_{0} = 90$ kOe. The decay of the echo amplitude after a $\pi/2- \tau -\pi$ pulse sequence was fit by the product of a time-independent Gaussian term, associated with nuclear dipole-dipole interaction, times an exponential decay described by $1/T_{2\rm{e}}$. The increase of $1/T_{2\rm{e}}$ observed above $T_{\rm{c}}$ evidences the onset of unconventional low-frequency fluctuations in the normal state, possibly involving nematic fluctuations. The arrow marks $T_{\rm{c}}$ in zero field.}\label{T2Ba122}
\end{figure}

While the studies of the nematic susceptibility have concentrated on 122 compounds owing to the possibility of growing large detwinned single crystals, an extensive study of the nematic transition under strain with NMR has not been carried out in other IBS. Remarkably, signatures of nematic fluctuations are present throughout the whole phase diagram of BaFe$_{2}$As$_{2}$, both in the electron \cite{Bos13,Bos16} and in the hole doped regime \cite{Mor19}. In particular, recently it has been pointed out \cite{Ish18} that in the hole-doped RbFe$_{2}$As$_{2}$ compound the nematic susceptibility derived from elastoresistivity measurements diverges around 35 K when the strain is applied along the diagonal of the Fe square-lattice unit cell, and does not diverge when the susceptibility is applied perpendicular to that direction, the orientation for which $\chi_{\rm{nem}}$ diverges in BaFe$_{2}$As$_{2}$ \cite{Chu12b}. The origin of this change of symmetry of the nematic order is currently subject of debate \cite{Bor19}. At the same temperature a peak in $^{75}$As NQR $1/T_{1}$ was detected (see Fig.~\ref{XYnemfig}). This peak indicates a progressive slowing down of the fluctuations on approaching the nematic transition. The peak is absent for $^{75}$As NMR $1/T_{1}$ experiments at $H = 70$ kOe on the same sample, pointing out either an extreme softening of the fluctuations to frequencies approaching the Larmor frequency or to a magnetic field induced reduction of these fluctuations. It is noticed that charge nematic fluctuations can cause a modulation of the EFG at $^{75}$As nuclei and give rise to an additional quadrupolar relaxation mechanism for $1/T_{1}$ \cite{Dio16}.

\begin{figure}
	\vspace{5.2cm} \includegraphics{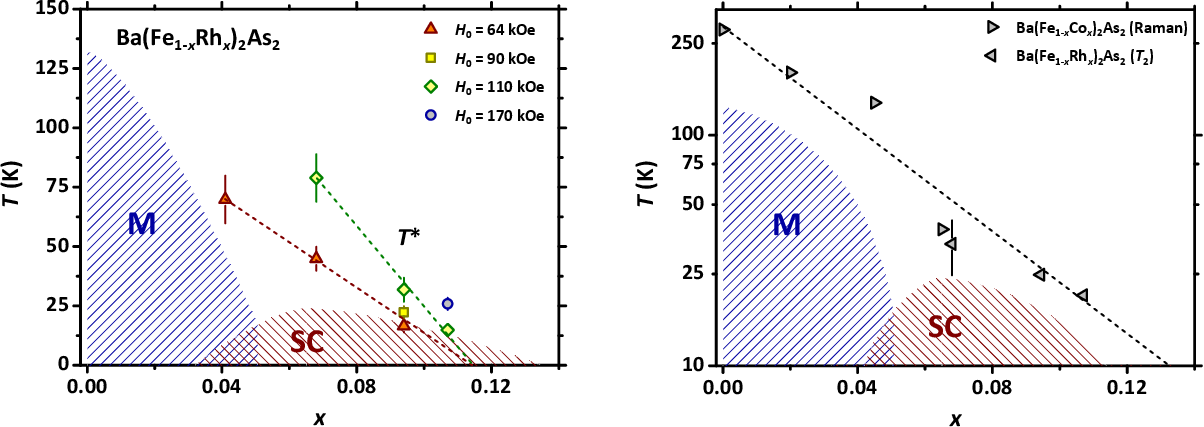} \caption{Left-hand panel: electronic phase diagram of Ba(Fe$_{1-x}$Rh$_{x}$)$_{2}$As$_{2}$ showing the characteristic temperature where low-frequency fluctuations start to cause an increase of $^{75}$As $1/T_{2\rm{e}}$ for different magnetic field intensities. The right-hand panel shows the behaviour of the energy barrier characterizing the temperature dependence of the fluctuations' correlation time, as derived from Raman susceptibility measurements in Ba(Fe$_{1-x}$Co$_{x}$)$_{2}$As$_{2}$ \cite{Gal13,Gal16} and from $1/T_{2}$ measurements \cite{Bos16} in Ba(Fe$_{1-x}$Rh$_{x}$)$_{2}$As$_{2}$.}\label{T2diag}
\end{figure}

It is interesting to notice that further signatures of very low-frequency fluctuations, possibly associated with nematic fluctuations, were detected also in the case of electron-doped BaFe$_{2}$As$_{2}$. In particular, in the normal state of Rh-doped BaFe$_{2}$As$_{2}$, displaying properties very similar to Co-doped BaFe$_{2}$As$_{2}$ \cite{Can10}, a clear evidence of fluctuations slowing down to frequencies below the MHz has been presented, both on the basis of $^{75}$As $1/T_{1}$ and of $1/T_{2}$ measurements \cite{Oh11,Bos13,Bos16}. In fact, the low-frequency dynamics lead to the appearance of an exponential component in the echo decay, characterized by a $1/T_{2}$ relaxation rate which increases already in the normal state (see Fig.~\ref{T2Ba122}). These fluctuations, which persist well into the overdoped regime (see Fig.~\ref{T2diag}), give rise to a peak in $1/T_{1}$ when the magnetic field is aligned perpendicular to the $c$-axis but which is absent when the magnetic field is along the $c$ axis (see Fig.~\ref{FigT1Ba122}), suggesting that the fluctuations give rise to a component of the local field along the $c$ axis. The comparison of the echo decay time $T_{2}$ derived both with a standard Hahn echo sequence with that derived from CPMG pulse sequence \cite{Sli90} provides further evidence for the presence of very low-frequency fluctuations in the normal state. In particular, one finds that the decay rate of the CPMG progressively increases with the delay between the CPMG pulses \cite{Bos16} which is a clear signature of dynamics with a characteristic time scale of the order of the delay between the pulses, namely around the tens of $\mu$s to ms. It should be noticed that these low-frequency fluctuations have been rarely detected in the cuprates but are clearly detected in the normal phase of vanadates which are prototypes of the $J_{1}-J_{2}$ model on a square lattice, where Ising nematic fluctuations are present \cite{Cha90,Car02}. Given the similarities in the magnetic structure of IBS and vanadates, it has been argued that these fluctuations are nematic ones \cite{Bos16}. Further support in this respect comes from the comparison of the characteristic barrier describing the activated slowing down of the fluctuations derived from $T_{2}$, $T_{1}$ and Raman scattering in electron-doped BaFe$_{2}$As$_{2}$ (see Fig.~\ref{T2diag}), where the latter technique has clearly evidenced electronic nematic fluctuations \cite{Gal13,Gal16}. It turns out that the characteristic correlation time probed by these different techniques can be fitted with an Arrhenius law characterized by an energy barrier which progressively decreases upon increasing the electron doping and which is quantitatively close for the same doping level (see Fig.~\ref{T2diag}). The presence of anomalous low-frequency fluctuations have also been detected in F-doped LaFeAsO and also here, from the study of $1/T_{1}$ at different magnetic fields, it was possible to evidence that the fluctuations slow down to frequencies of the order of the Larmor frequency \cite{Ham13}.

\subsection{Charge order}\label{Chargesect}

One of the most debated issues in the field of high-$T_{\rm{c}}$ superconducting cuprates is the role of the charge order which has been extensively probed by X-ray scattering techniques over the last decade \cite{Ghi12,Huc14}. Are charge density wave excitations directly involved in the pairing mechanism, are they promoting indirectly superconductivity or the CDW (charge density wave) order is actually competing with superconductivity? The first evidence for the presence of a CDW order in the phase diagram of the cuprates came from NMR \cite{Wu11} and the fingerprint was the appearance of a splitting in the $^{63}$Cu NMR spectra in two slightly resolved bumps. The splitting is due to a modulation of the EFG at the nucleus and indicates a periodicity of the charge distribution which is no longer described by the crystallographic unit cell. In several cuprates the splitting of the NMR spectra cannot be resolved and one actually detects a broadening in the CDW phase \cite{Wu15}. In a Pt pnictide superconductor, SrPt$_{2}$As$_{2}$, which share many similarities with the iron-based superconductors, even two CDW phases were clearly detected above the superconducting transition temperature by means of $^{75}$As NQR \cite{Kaw15}, causing a remarkable broadening of the NQR and NMR spectra.
\begin{figure}
	\vspace{10.7cm} \includegraphics{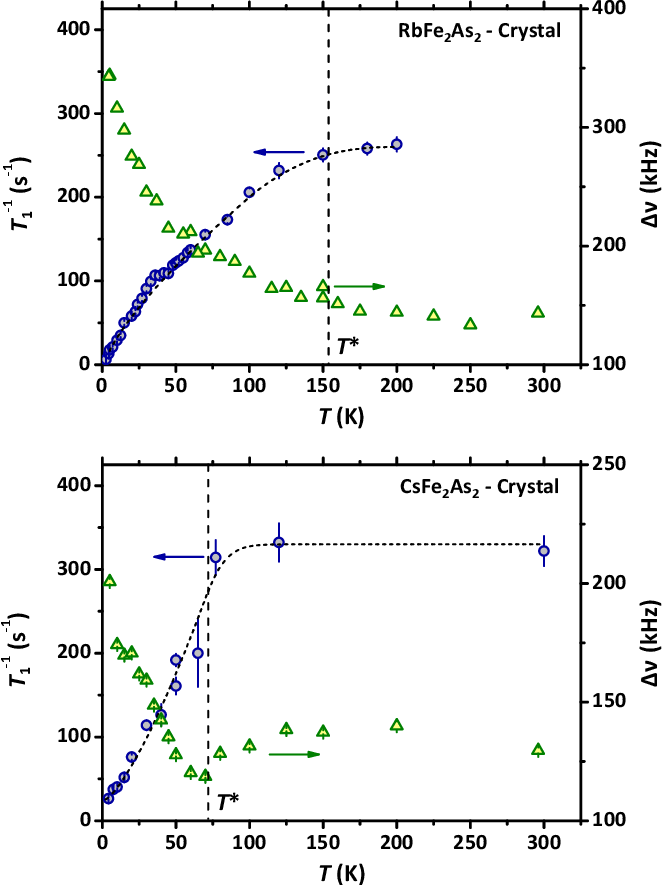} \caption{Temperature dependence of $^{75}$As NQR $1/T_{1}$ and linewidth in RbFe$_{2}$As$_{2}$ (top panel) and in CsFe$_{2}$As$_{2}$ (bottom panel) single crystals. The vertical lines mark the temperature $T^{*}$ where a crossover in the electronic structure is present, likely involving a charge order. The other lines are guides to the eye. }\label{Cs122T1}
\end{figure}

What about the iron-based superconductors? As remarked previously, in 1111 compounds $^{75}$As NQR spectra are characterized by double peaks for certain fluorine doping levels which evidence a nanoscopically-modulated charge distribution, possibly pinned by disorder \cite{Lan10}. This inhomogeneity is already present at room temperature. Studies of the NQR spectra above room temperature may suffer from the quenching of that static charge distribution due to the ionic motions of the dopants and hence it is not clear if the narrowing of the NQR spectra originates from the ionic motions or from a reduction of a charge order parameter. In fact, $^{75}$As NQR measurements in LaFeAsO$_{1-x}$F$_{x}$ up to $550$ K show a thermal history associated with fluorine diffusion and so far no clear conclusion was drawn about the disappearance of an intrinsic charge modulation. 

\begin{figure}
	\vspace{9.3cm} \includegraphics{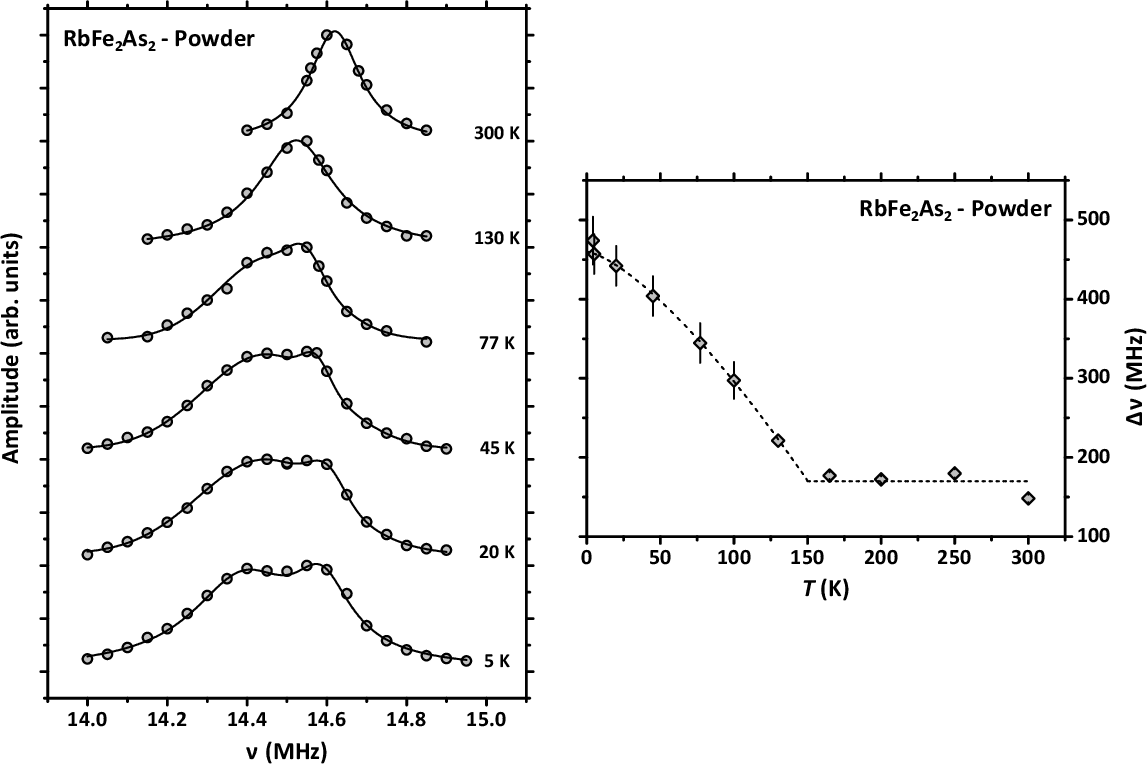} \caption{Left-hand panel: $^{75}$As NQR spectrum in RbFe$_{2}$As$_{2}$ powders at different temperatures. Right-hand panel: temperature dependence of the corresponding full width at half intensity, evidencing a clear increase below $T^{*} \simeq 140$ K. This quantity, proportional to the amplitude of the EFG modulation, provides the temperature dependence of the order parameter of the charge ordered state.}\label{NQRCharge}
\end{figure}
In the cuprates the charge order phase gets enhanced in the underdoped regime where the effects of the strong electronic correlations are more pronounced. Accordingly, it is convenient to search for an analogous phase in the alkali-doped 122 compounds where the substitution of Ba with an alkali ion causes a reduction in the number of e$^{-}$/Fe to 5.5, approaching the half-filling condition where eventually a Mott insulating phase could arise. $^{75}$As NQR experiments in RbFe$_{2}$As$_{2}$ and in CsFe$_{2}$As$_{2}$ showed that $^{75}$As $1/T_{1}$ shows a crossover below a characteristic temperature $T^{*}$ from a high-temperature $T$-independent behaviour to a low-temperature power-law behaviour (see Fig.~\ref{Cs122T1}) \cite{Mor19,Wu16,Civ16}. This crossover from a high-$T$ trend, which is more characteristic of a phase with nearly localized electrons, to a low-$T$ phase with a $1/T_{1}$ behaviour more characteristic of delocalized electrons suggests that around $T^{*}$ a change in the Fermi surface is taking place. Remarkably, at the same temperature a broadening of the NQR spectra is detected both in CsFe$_{2}$As$_{2}$ and in RbFe$_{2}$As$_{2}$ single crystals (see Fig.~\ref{Cs122T1}). In RbFe$_{2}$As$_{2}$ powders, where the broadening is more pronounced, one detects the emergence of two bumps in the NQR spectra (see Fig.~\ref{NQRCharge}) which indicate the presence of at least two inequivalent $^{75}$As sites, namely that the periodicity of the charge distribution is no longer defined by the crystal lattice step. The NQR line broadening is proportional to the amplitude of the charge modulation, corresponding to the order parameter and its temperature dependence is shown in Fig.~\ref{NQRCharge}. 

In spite of the presence of a charge order in the IBSs, akin to the one detected in the cuprates, the nature of the charge order is still unclear. Namely, it is not established if there is a CDW, an orbital order or another type of inhomogeneous charge distribution. Indeed also the mechanism driving that inhomogeneous charge distribution could differ between the two families of superconductors. In particular, in the IBS the pronounced Hund coupling $J$ could drive a charge unbalance. In fact, Isidori \textit{et al.} \cite{Isi19} pointed out that when $J/U$ is sufficiently large, with $U$ Hubbard Coulomb repulsion, the tendency to promote high-spin configurations would favour a transfer of charge between different sites and generate a charge disproportionation. In that case, as pointed out by Moroni \textit{et al.} \cite{Mor19}, different local charge configurations can be realized and accordingly a inhomogeneous EFG distribution is present, leading to the broadening of the NQR spectra. An analogous charge distribution was suggested by Wang \textit{et al.} \cite{Wan16} to explain the splitting of the $^{75}$As NMR line in KFe$_{2}$As$_{2}$ at high pressure. It is envisaged that these results will trigger further research with other techniques \cite{Mar17} to provide other evidences for a charge order.
\begin{figure}
	\vspace{6.5cm} \includegraphics{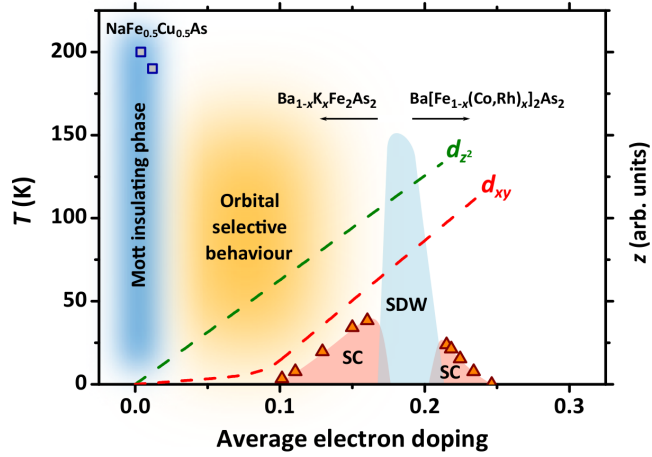} \caption{Schematic phase diagram of the 122 family of IBSs as a function of the average electron doping per Fe orbital, referred to the half-filling condition. Namely if $n_{\rm{e}}$ is the number of electrons per Fe, the horizontal scale is given by $x = (n_{\rm{e}}-5)/5$. The superconducting (SC) and the magnetically ordered SDW phases around 0.18 are shown. For $x\simeq 0$ a magnetically ordered Mott-Hubbard insulating phase is present, as recently reported for NaFe$_{1-y}$Cu$_{y}$As \cite{Xin19,Xin19b}. Other pnictides at half band filling are BaMn$_{2}$As$_{2}$ \cite{Joh11} and BaCrFeAs$_{2}$ \cite{Fil17}. In the $0\leq x \leq 0.2$ region a marked orbital selective behaviour is expected \cite{DeM09,DeM14,DeM15} as a consequence of the different quasiparticle weight $z\sim 1/m^{*}$, with $m^{*}$ the effective mass, for the different bands crossing the Fermi level. The dashed lines show the qualitative behaviour of $z$ for the $d_{z^{2}}$ and $d_{xy}$ bands, evidencing that for the latter one $z$ is vanishing when it is still sizeable for the other band. Accordingly an orbital selective Mott transition is expected.}\label{figorbsel}
\end{figure}

\subsection{Orbital selectivity}\label{Sectorb}

Within an orbitally-selective scenario, if a nucleus is coupled with different electron bands then $1/T_{1}$ would be the sum of independent contributions from each of these bands. A first clear evidence of such a behaviour came from $^{75}$As NMR study of the magnetically ordered LaFeAsO \cite{Nak08} where the nuclear spin-lattice relaxation can be decomposed as the sum of a Korringa-like linear trend of $1/T_{1}$ with temperature, characteristic of a metal, superimposed to a behaviour which is more characteristic of nearly-localized electrons with a progressive increase of the spin correlations upon cooling and eventually diverging at the magnetic ordering temperature.

On the other hand, one would expect that different nuclei could be more or less coupled with different bands depending on the hybridization of the orbitals of the atoms to which they belong to and, accordingly, show a behaviour of $1/T_{1}$ which is more characteristic of strongly correlated or weakly correlated metals. Since the orbital selective behaviour is expected to become more pronounced with increasing electronic correlations it is likely that a more clear signature of this behaviour is evidenced in the alkali-doped BaFe$_{2}$As$_{2}$, with an average electron doping of 5.5 e$^{-}$/Fe. Indeed, Zhao \textit{et al.} \cite{Zha18} presented a clear evidence in CsFe$_{2}$As$_{2}$ for a different temperature dependence for $^{133}$Cs and $^{75}$As $1/T_{1}$ as well as for their paramagnetic NMR shift. While at low temperature the temperature dependence of these quantities is the same for both nuclear species, above a temperature $T^{*}$, the one where a charge order is suggested to develop, one notices a flattening of $^{75}$As $1/T_{1}$ and a monotonous increase of $^{133}$Cs $1/T_{1}$. The former is more characteristic of nearly localized electrons while the latter of weakly correlated ones. Remarkably, in RbFe$_{2}$As$_{2}$ and in KFe$_{2}$As$_{2}$ $^{75}$As the spin-lattice relaxation rate is characterized by two components at low temperature \cite{Civ16,Wie18}, again one more characteristic of nearly localized electrons while the other of a Fermi liquid, suggesting a possible tendency towards a nanoscopic phase separation in more localized and more metallic regions.

\section{Probing iron based superconductivity with impurities}

\begin{figure}
	\vspace{6.5cm} \includegraphics{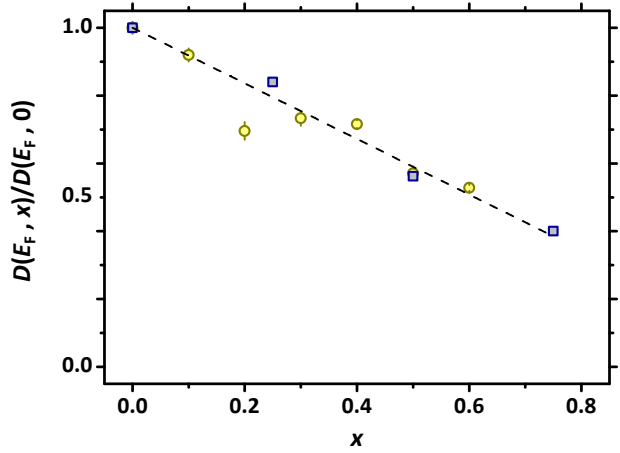} \caption{Doping dependence of the density of states at the Fermi level in LaFe$_{1-x}$Ru$_{x}$AsO as deduced from the doping dependence of $^{139}$La $1/T_{1}$ around room temperature (squares) and from band structure calculations (circles) \cite{Bon12}.}\label{DOSRu}
\end{figure}
The understanding of the effect of impurities in superconductors is relevant for many aspects \cite{All09,Sur17}. First of all it allows to probe the stability of superconductivity against unwanted defects which could be present during the manufacturing of superconducting devices. They can act as pinning centre and enhance the critical currents and the performance of superconductors at high magnetic fields. Moreover, the effect of impurities on $T_{\rm{c}}$ and on the superconducting order parameter allows to distinguish among the possible gap symmetries \cite{Vav11,Fer12b} and allows one to investigate the local susceptibilities and to unravel quantum critical points.

\begin{figure}
	\vspace{14.7cm} \includegraphics{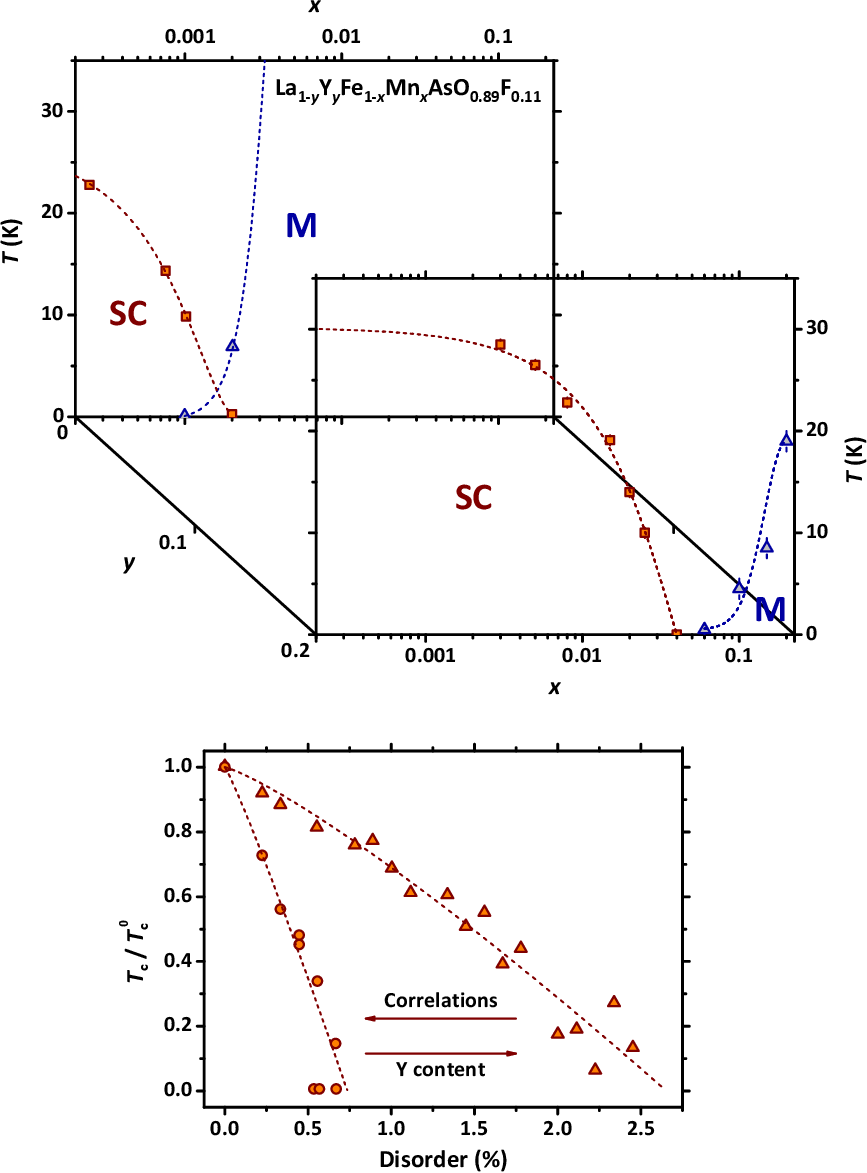} \caption{Top panel: phase diagram of optimally electron doped LaFeAsO$_{0.89}$F$_{0.11}$ as a function of the amount of paramagnetic impurities for two different Y contents (0 and 0.2). The effect of Y is to enhance the $t/U$ ratio, with $t$ the hopping integral, and accordingly decrease the effect of electronic correlations which markedly suppress $T_{\rm{c}}$, as shown by the theoretical calculations reported in the bottom panel \cite{Gas16,Mor17b}.}\label{MnPhDiag}
\end{figure}
The decrease of $T_{\rm{c}}$ upon doping with impurities can be significantly different if one considers the substitution of Fe with a diamagnetic atom as Ru \cite{Tro10,San11} or with a paramagnetic one as Mn \cite{Ham14}. In 1111 compounds it has been observed that Ru causes a very slight decrease of $T_{\rm{c}}$ with doping, eventually vanishing only at doping levels as high as $\simeq 60$ \% of Ru \cite{San13}. Such a tiny effect can be understood only if one considers the formation of Ru clusters (dimers, trimers, etc\ldots) which tend to enhance the local density of states for $d_{z^{2}}$ orbitals and stabilize both the superconducting and the magnetic phases \cite{Gas16}. In fact, also in the parent magnetic compound LaFeAsO the effect of Ru doping on the N\'eel temperature is very weak and also here the ordering temperature vanishes around $60$ \% of Ru doping \cite{Bon12}. It is remarkable that this value is close to the percolation threshold for a $J_{1}-J_{2}$ model on a square lattice \cite{Pap05} which has been often used as a starting point to analyse the magnetic properties of the parent compounds of IBSs. In fact, one observes an initial suppression rate for the magnetic ordering temperature $(1/T_{\rm{N}}(x=0))(dT_{\rm{N}}/dx)$ which is very close to that of spin-diluted Li$_{2}$VOSiO$_{4}$ \cite{Pap05}, a prototype of the $J_{1}-J_{2}$ model on a square lattice \cite{Mel00}. Ru is also found to yield a progressive decrease of $1/T_{1}$ at high temperature which can be associated with a decrease in the density of states (see Fig.~\ref{DOSRu}) induced by the broadening of the bands caused by the more extended Ru $4d$ orbitals \cite{Tro10}.

On the other hand, the effect of Mn doping on the optimally F-doped LaFeAsO superconductor has a dramatic effect on the superconducting phase: it is enough to introduce a tiny Mn doping, as low as 0.2 \% to fully suppress the superconducting transition temperature from 26 K to zero \cite{Ham14,Sat10,Sat12}. The disruption of superconductivity is also accompanied by charge localization \cite{Sat10,Kap18}, suggesting that Mn doping significantly modifies the electronic properties \cite{Mar19}. Such a dramatic effect cannot be justified within an independent impurity model but points towards significant correlations among the Mn impurities (see Fig.~\ref{MnPhDiag}) \cite{Gas16,Mor17b} or, in other terms, to a particularly enhanced non-local spin susceptibility, as it is expected in the proximity to a quantum critical point. 

\begin{figure}
	\vspace{5.5cm} \includegraphics{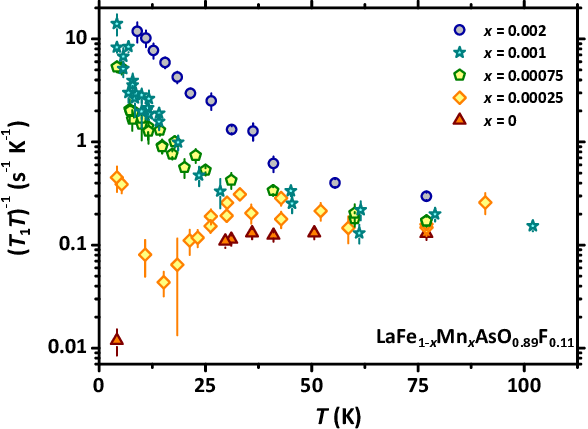} \includegraphics{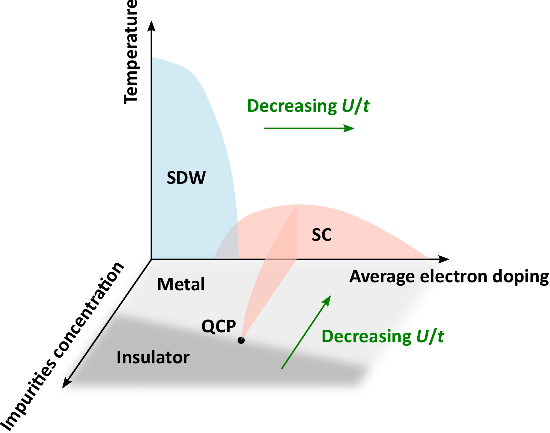} \caption{Left-hand panel: temperature dependence of $^{75}$As NQR $1/T_{1}$ in optimally electron doped LaFe$_{1-x}$Mn$_{x}$AsO$_{0.89}$F$_{0.11}$ for different Mn contents. A neat divergence is noticed for $x\simeq 0.002$ where a quantum critical point is present and $T_{\rm{c}}$ vanishes. Right-hand panel: schematic phase diagram representing the effect of Mn impurities in the phase diagram of 1111 compounds.}\label{T1AsMn}
\end{figure}

Indeed, a careful study of the temperature and Mn-dependence of $^{75}$As $1/T_{1}$ showed that as the Mn content increases towards $x_{\rm{c}}\simeq 0.002$, the value leading to the suppression of $T_{\rm{c}}$, $1/T_{1}T$ increases at low temperature and eventually diverges at $x_{\rm{c}}$ for $T \rightarrow 0$ (see Fig.~\ref{T1AsMn}) \cite{Ham14}. These results were analyzed in the framework of the self-consistent renormalization theory \cite{Ish96,Ish98} which has been often used to describe strongly correlated electron systems close to a quantum critical point. For a quasi-2D antiferromagnetically correlated system one finds:
\begin{equation}
\frac{1}{T_{1}}= \frac{\hbar\gamma^{2} \mathcal{A}^{2}}{4\pi}\left( \frac{T}{T_{0}}\right) \chi(Q_{\rm{AF}})\,\, ,
\end{equation}
with $T_{0}$ a temperature describing the characteristic frequency of the spin fluctuations \cite{Ish96,Ish98}, $\mathcal{A}$ the hyperfine coupling and $\chi(Q_{\rm{AF}})$ the static spin susceptibility at the antiferromagnetic wavevector. By expressing this latter quantity in terms of the in-plane spin correlation length $\xi$ \cite{Ham14,Ish96} one can establish a direct correspondence between $T_{1}$ and $\xi$ 
\begin{equation}
\frac{1}{T_{1}}= \frac{\hbar\gamma^{2} A^{2} S(S+1)}{4\pi 3 k_{\rm{B}} T_{0}} \frac{4\pi \xi^{2}}{\ln[4\pi \xi^{2} + 1]}\,\, ,
\end{equation} 
with $S$ the electron spin. From the temperature dependence of $1/T_{1}$ at $x_{\rm{c}} = 0.002$ of Mn one derives $\xi\sim 1/\sqrt{T}$, the temperature dependence of the correlation length expected for a 2D correlated antiferromagnetic metal close to a quantum critical point (see Fig.~\ref{T1AsMn}). This divergence marks the quantum phase transition from a superconducting to a magnetic ground-state. The nature of the magnetic ground-state has been discussed for years until recently Moroni \textit{et al.} \cite{Mor17b} found from zero-field $^{75}$As NMR measurements that the magnetic order is stripe, i.e., characterized by $Q = \left(\frac{\pi}{a},0\right)$ or $\left(0,\frac{\pi}{a}\right)$ magnetic wavevector, as the parent LaFeAsO compound. Remarkably, the recovery of the magnetic order is followed by a recovery of the nematic transition which leads to a change in $^{75}$As quadrupolar frequency \cite{Mor17b}, suggesting that nematicity is driven by spin correlations. It is also noticed that the low-frequency fluctuations giving rise to a peak in $1/T_{1}$ above $T_{\rm{N}}$ get enhanced by Mn-doping \cite{Ham15}. This enhancement can be associated with an increase in the spin polarization around Mn impurities. 

\begin{figure}
	\vspace{6.5cm} \includegraphics{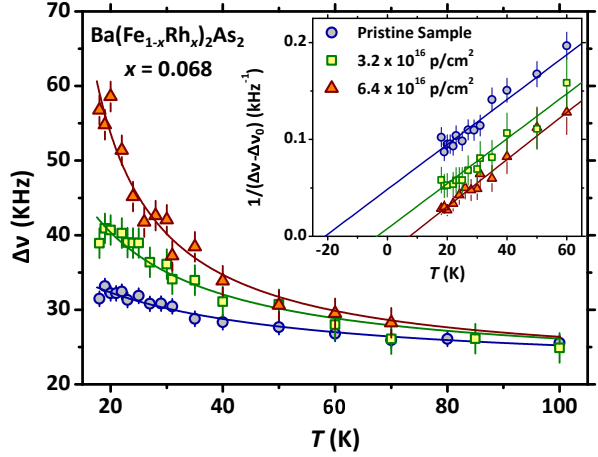} \caption{Effect of different doses of proton irradiation on the $^{75}$As NMR linewidth in Ba(Fe$_{0.932}$Rh$_{0.068}$)$_{2}$As$_{2}$ for a $70$ kOe magnetic field along the $c$ axes \cite{Mor17c}. The inverse of the linewidth is reported in the inset in order to evidence the change of sign of the Curie-Weiss temperature, with $\Delta \nu_{0}$ a temperature independent term.}\label{Dnuirr}
\end{figure}

Defects can also be introduced by irradiating the materials with electrons, heavy ions or protons. In the latter case a rather uniform distribution of point defects can be achieved over a sample thickness of about 50 $\mu$m \cite{Nak10}. NMR experiments on optimally doped and overdoped Ba(Fe$_{1-x}$Rh$_{x}$)$_{2}$As$_{2}$ for different radiation doses clearly show a broadening of $^{75}$As line upon increasing the number of defects (see Fig.~\ref{Dnuirr}) \cite{Mor17c}, as it is expected. The temperature dependence of the linewidth follows a Curie-Weiss law with a non-zero Curie-Weiss temperature, as it is expected in the presence of spin correlations. Remarkably the Curie-Weiss temperature changes sign by increasing the radiation dose, indicating a crossover from antiferromagnetic to ferromagnetic spin correlations for the highest dose (see Fig.~\ref{Dnuirr}) \cite{Mor17c}. It is noticed that while there is a clear effect of proton irradiation on the NMR linewidth, the proton doses used in these experiments, leading to a distance between the defects of the order of the coherence length, cause a very tiny reduction of the superconducting transition temperature as well as of $1/T_{1}$. This suggests that while the static susceptibility can be significantly affected by the defects the low-energy spin excitations as well as $T_{\rm{c}}$ are not sizeably changed, coherently with a spin driven pairing mechanism.

\section{The superconducting state}

\subsection{The superconducting order parameter}

NMR has proven to be one of the most powerful tools to unravel the microscopic mechanisms leading to superconductivity \cite{Tin96,Smi06}. From the temperature dependence of $1/T_{1}$ and of the paramagnetic shift $\Delta K$ one can probe the symmetry of the superconducting gap in the reciprocal space and, accordingly, put some constraints on the potential pairing mechanism -- e.g., phononic or due to spin excitations. In the conventional type I superconductors, the study of the superconducting state requires non-conventional approaches as the field cycling in order to avoid an unwanted suppression of the superconducting phase by the magnetic field needed to perform NMR experiments. On the other hand, in type II superconductors such as cuprates and IBSs the upper critical fields are quite high, in some cases exceeding $10^{6}$ Oe. Then, over a wide portion of the magnetic field-temperature ($H - T$) phase diagram a FLs (flux lines) lattice is present and standard NMR experiments can be performed without the need of field cycling, even if a screening of the radiofrequency and hence a partial reduction of the NMR signal can be present. 

Thanks to the hyperfine coupling of the nuclei with the electron spins, the measurement of the paramagnetic shift is basically the only direct way to probe the spin-state of the Cooper pair and to clarify whether singlet or triplet superconductivity is realized. In the first case a decrease in the absolute value of $\Delta K$ is observed below $T_{\rm{c}}$ \cite{Kit11}, at variance with the case of supercondutors with a triplet spin ground-state. In all IBSs a spin-singlet state is observed and hence the orbital wave function must be characterized by an even symmetry (e.g. $s$-wave). The temperature dependence of the paramagnetic shift in a conventional $s$-wave supercondutor, characterized by a uniform opening of the gap in the reciprocal space, causes an exponential decrease of the paramagnetic shift at low temperature. On the other hand, different symmetries characterized by nodes in the gap or by the presence of more than one gap \cite{Kla11} give rise to a power-law decrease or even to more complex temperature dependences for $\Delta K$. Several studies of IBSs have pointed out that the paramagnetic shift follows a power-law at low temperature which could be consistent with an $s_{\pm}$ symmetry of the order parameter, associated with interband coupling involving electrons and holes \cite{Nis10,Yan12}.
\begin{figure}
	\vspace{7.0cm} \includegraphics{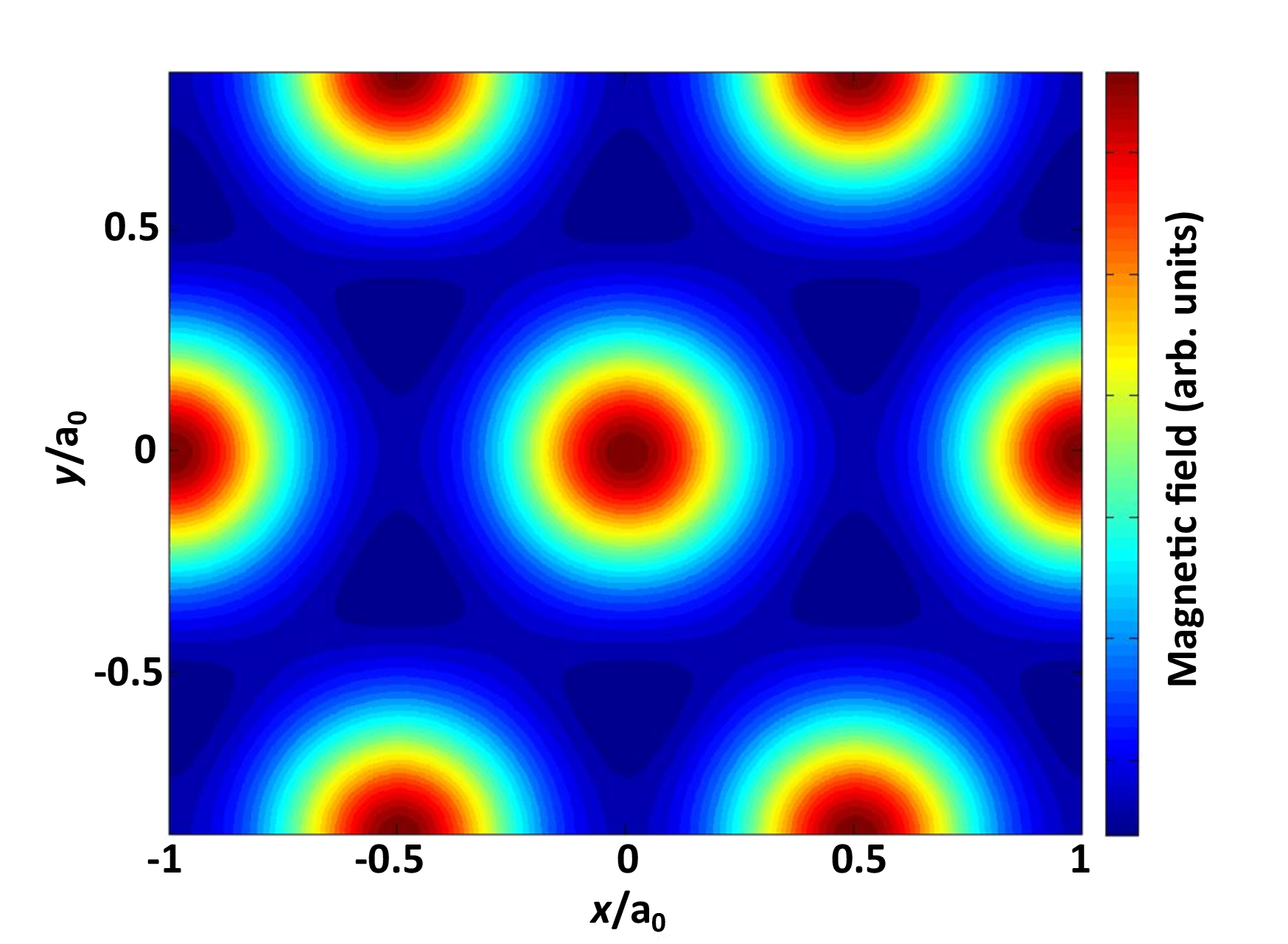} \caption{Contour plot for the magnetic field distribution in a flux lines lattice characterized by a step $a_{0}$, for $\bm{H}_{0} // z$. The field is maximum (red) at the core of the flux lines and minimum (blue) at the centre of the triangular unit cell. The most probable field is the one at the saddle point between two flux lines.}\label{FLLdist}
\end{figure}

The temperature dependence of $1/T_{1}$ allows to investigate the symmetry of the superconducting gap as well. In the presence of a wavevector-dependent gap $\Delta(\bm{k})$, one can write $1/T_{1}$ in terms of $\Delta(\theta,\phi)$ \cite{Muk10}
\begin{equation}\label{T1SC}
	\frac{(T_{1}T)^{-1}}{(T_{1}T)_{T_{\rm{c}}}^{-1}}= \frac{2}{k_{\rm{B}}T_{\rm{c}}}\int_{0}^{\infty} \left[N(E)^{2} + \alpha_{\rm{c}} M(E)^{2}\right]f(E)[1-f(E)] dE
\end{equation}
with $f(E)$ Fermi-Dirac distribution function, $\alpha_{\rm{c}}$ a phenomenological factor weighting the interband contribution,
\begin{eqnarray}
N(E)= \int_{\rm{BZ}} \frac{E}{\sqrt{E^{2}-|\Delta(\bm{k})|^{2}}} d\bm{k}
\end{eqnarray}
the quasiparticle density of states and
\begin{eqnarray}
M(E)= \int_{\rm{BZ}} \frac{\Delta(\bm{k})}{\sqrt{E^{2}-|\Delta(\bm{k})|^{2}}} d\bm{k}\,\, .
\end{eqnarray}
$M(E)$ is the coherence term in the density of states. Generally, the temperature dependence of $1/T_{1}$ for $T\rightarrow 0$ in IBSs is characterized by a power-law behaviour which indicates either the presence of nodes and/or of more than one superconducting gap. The overall behaviour of $1/T_{1}$ is also consistent with an $s^{\pm}$ symmetry of the superconducting gap. It is remarked that in F-doped LaFeAsO no Hebel-Slichter peak \cite{Heb59} is observed below $T_{\rm{c}}$ up to optimal doping, as in most families of IBSs -- i.e., the $\alpha_{\rm{c}}$ term in Eq.~\ref{T1SC} is negligible. However, a coherence peak has been reported in SrPtAs \cite{Mat14} and in the heavily overdoped LaFeAsO$_{1-x}$F$_{x}$ \cite{Muk10}. This suggests a reduction of the interband scattering between electron and hole Fermi surface in these latter compounds, leading to a sizeable increase in $\alpha_{\rm{c}}$ term -- coherently with some theoretical predictions \cite{Yam12}.

\subsection{Probing the lattice of flux lines with nuclear spins}\label{Sectflux}

Nuclear spins act as nanoscopic Hall sensors of the magnetic field distribution within the FLs lattice (see Fig.~\ref{FLLdist}) and of its dynamics. Since in a magnetic field of a few tens of kOe the FLs lattice step is of the order of hundreds of \AA, i.e. much larger than the crystal lattice step, the nuclei allow to map with a sufficiently good resolution the field distribution within it. Although the penetration of the static magnetic field takes place over a wide range of the $H-T$ diagram, the penetration of the radio-frequency field is more subtle and depends very much on the mobility of FLs. In fact, the radio-frequency penetration depth depends not only on the London penetration depth but also on the mobility of FLs through the Campbell penetration length \cite{Cof92,Cof92b}. Accordingly, when the mobility of FLs is high the radio-frequency field can shake the FLs and be transmitted throughout the sample volume. Otherwise, when the FLs lattice is static the penetration depth is limited by the London penetration depth and, depending on the sample volume and shape, the radio-frequency field can significantly decrease in the superconductor and hence the NMR signal is reduced \cite{Car93}. If the radio-frequency power is sufficiently high and the FLs are static the application of the radio-frequency pulses may induce the sample vibration and accordingly enhance the magnetoacoustic ringing arising after the pulses, leading to an unwanted increase of the deadtime in the data acquisition \cite{Bos14}.

\begin{figure}
	\vspace{5.2cm} \includegraphics{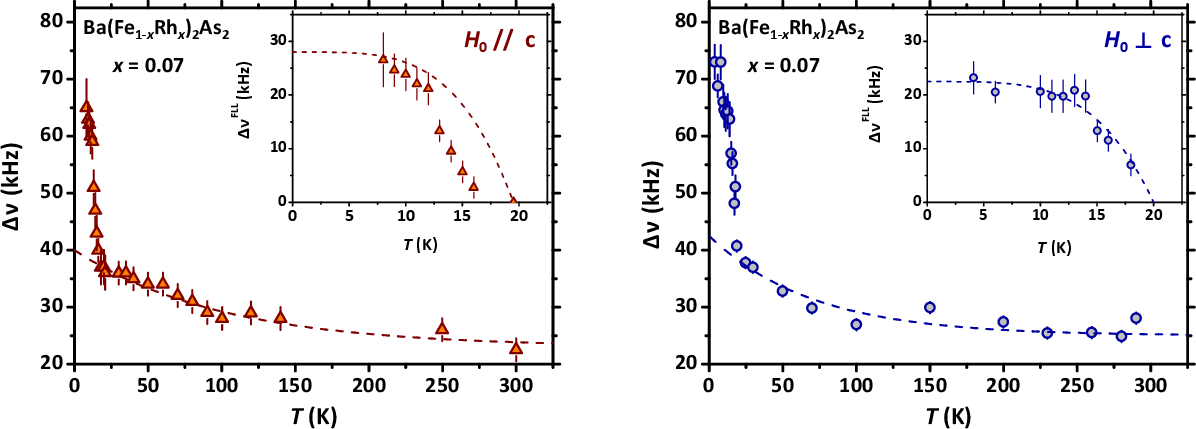} \caption{Temperature dependence of the $^{75}$As NMR linewidth in Ba(Fe$_{0.93}$Rh$_{0.07}$)$_{2}$As$_{2}$ for $\bm{H}_{0} // c$ and $\bm{H}_{0} \perp c$ (left- and right-hand panels, respectively), evidencing a clear increase at $T_{\rm{c}}$. The dashed lines track the temperature dependence in the normal state. Insets: temperature dependence after subtracting the normal state contribution to the broadening. The dashed lines show the behaviour expected below $T_{\rm{c}}$ according to a two-fluid model.}\label{DnuFLL}
\end{figure}
The field distribution generated by the FLs lattice causes an additional broadening of the NMR spectrum which is directly related to the second moment of the distribution and hence to the penetration depth $\lambda$ \cite{Bra86,Bra88}. Thus, if for simplicity one fits the NMR spectrum with a Gaussian function, the linewidth at half intensity
\begin{equation}\label{eqdnufll}
	\Delta\nu\simeq \frac{2 \gamma\sqrt{2 \ln 2}}{2\pi} \sqrt{\langle \Delta B^{2} \rangle}\simeq 2.36 \gamma \frac{k_{\rm{FL}}\Phi_{0}}{\lambda^{2}}
\end{equation}
with $\Phi_{0}$ the flux quantum and $k_{\rm{FL}}$ a number which depends on the unit cell of the FLs lattice \cite{Bra86}. From the additional broadening of $^{75}$As NMR line in Ba(Fe$_{0.93}$Rh$_{0.07}$)$_{2}$As$_{2}$ below $T_{\rm{c}}$, for $\bm{H}_{0} // c$, (see Fig.~\ref{DnuFLL}) one derives $\lambda\simeq 240$ nm \cite{Bos12}, a value which is slightly larger than what one would deduce from other techniques not suffering from issues related to the penetration of radio-frequency. Remarkably, the temperature dependence of $\Delta\nu$ is not the one expected according to a simple two-fluid model, $\Delta\nu\propto [1-(T/T_{\rm{c}})^{4}]$, but some narrowing with respect to this trend is noticed. The narrowing could be associated with the symmetry of the superconducting order parameter, with the presence of more than one gap, or even with the dynamics of the FLs causing a motional narrowing of the NMR spectrum \cite{Rig98}.

\begin{figure}
	\vspace{6.5cm} \includegraphics{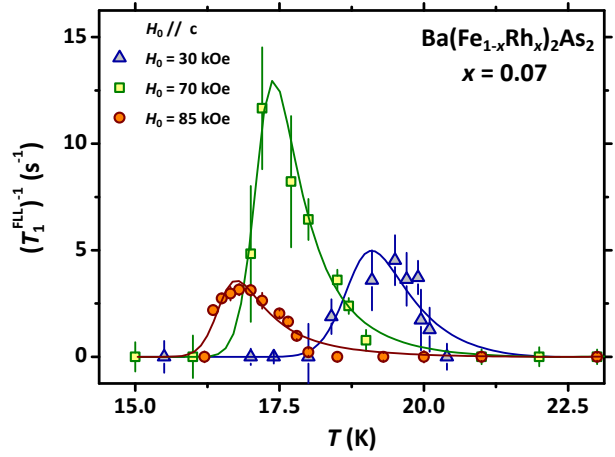} \caption{Temperature dependence of the dynamical contribution of FLs to $^{75}$As NMR $1/T_{1}$ in Ba(Fe$_{0.93}$Rh$_{0.07}$)$_{2}$As$_{2}$ for $\bm{H}_{0} // c$ at different values of magnetic field. A peak is observed when the characteristic frequency for the FLs motions matches the nuclear resonance frequency. As the field increases, the peak shifts to lower temperatures -- indicating faster dynamical processes at higher fields.}\label{T1FLL}
\end{figure}
While it is difficult to distinguish among these three mechanisms, more direct information on the dynamics can be derived from the temperature dependence of $1/T_{1}$. In particular, on cooling below $T_{\rm{c}}$ one expects a progressive slowing down of the dynamics of the FLs and when the characteristic frequency of the motions approaches the nuclear Larmor frequency a peak in $1/T_{1}$ should develop (see Sect.~\ref{SectT1T2}). The intensity of that peak depends \cite{Rig98} on the amplitude of the fluctuations, which in turn depends on the mean square displacement of the flux lines $\langle u^{2} \rangle$ from their equilibrium position, on the penetration and coherence lengths. Assuming nearly two-dimensional dynamics of the FLs in the $ab$ plane, namely that vortices are weakly coupled along the $c$ axes and characterized by a correlation time $\tau_{\rm{c}}(T,H)$, one finds
\begin{equation}\label{EqT1FLL}
	\frac{1}{T_{1}}= \frac{\gamma^{2}}{2} \frac{\Phi_{0}^{2} s^{2}}{4\pi\lambda_{\rm{c}}^{4}} \langle u^{2} \rangle \frac{\tau_{\rm{c}}(T,H)}{\xi_{ab}^{2} l_{\rm{e}}^{2} \sqrt{3}} \ln\left[\frac{\tau_{\rm{c}}(T,H)^{-2}+ \omega_{\rm{L}}^{2}}{\omega_{\rm{L}}^{2}}\right] \,\,\, .
\end{equation}
Bossoni \textit{et al.} \cite{Bos13b} were able to reproduce the temperature dependence of $^{75}$As $1/T_{1}$ in Rh-doped BaFe$_{2}$As$_{2}$ below $T_{\rm{c}}$ (see Fig.~\ref{T1FLL}) with Eq.~\ref{EqT1FLL} by assuming a temperature dependence of the correlation time described by the Vogel-Tammann-Fulcher (VTF) law
\begin{equation}
	\tau_{\rm{c}}(T,H) = \tau_{0} \exp\left[\frac{U(H)}{T-T_{0}(H)}\right]
\end{equation}
which is characteristic of glassy systems. Here, $U(H)$ is a characteristic energy barrier and $T_{0}$ the freezing temperature at which the correlation time diverges.

\begin{figure}
	\vspace{6.5cm} \includegraphics{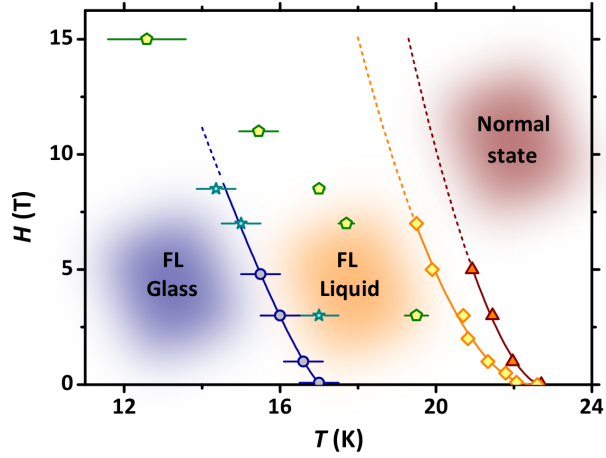} \caption{$H-T$ phase diagram for FLs in Ba(Fe$_{0.93}$Rh$_{0.07}$)$_{2}$As$_{2}$ for $\bm{H}_{0} // c$. The triangles show the upper critical field, while the diamonds mark the irreversibility temperature derived from the detuning temperature of the NMR probe. The pentagons show the temperature at which the peak in $1/T_{1}$ is detected (see Fig.~\ref{T1FLL}). The circles show the VFT temperature $T_{0}(H)$ derived from ac susceptibility measurements, while the stars the one derived from $1/T_{1}$ results \cite{Bos13b}.}\label{FLLPhDiag}
\end{figure}
Remarkably, a temperature dependence of the correlation time compatible with the VTF law was derived in the same compound also from $ac$ susceptibility measurements \cite{Bos13b} which, at variance with $1/T_{1}$, is sensitive only to fluctuations at $q \simeq 0$. From the combination of $1/T_{1}$ and susceptibility measurements, the $H-T$ phase diagram for the optimally electron doped Ba(Fe$_{0.97}$Rh$_{0.07}$)$_{2}$As$_{2}$ superconductor was derived, where a crossover from a high-temperature/high-field liquid phase to a low-temperature/low-field glassy phase is evidenced (see Fig.~\ref{FLLPhDiag}). In this context, it should be remarked that a different framework was discussed for Co-doped BaFe$_{2}$As$_{2}$ based on $ac$ susceptibility measurements \cite{Pra13b}. In particular, in this latter compound a vortex-glass phase transition could be detected based on the observation of a critical power-law dependence for the correlation time.

\section{Summarizing Remarks}

The weak perturbation approach used in NMR and even more in NQR, where no magnetic field is needed, combined with the local nature of the nuclear probe make these techniques rather unique and powerful for the understanding of the properties of superconductors. In this review we have presented some of the most recent achievements obtained by these technique in iron-based superconductors. In the superconducting state we have shown how NMR allows to detect the even $s^{\pm}$ symmetry of the superconducting order parameter and remains a rather unique approach to detect the related singlet spin state of the Cooper pair. Below $T_{\rm{c}}$ the nuclei allow also to probe the local field fluctuations induced by the flux lines dynamics and the combined investigation with $1/T_{1}$, $1/T_{2}$ and NMR spectroscopy allows to estimate the London penetration depth, the progressive slowing down of the dynamics on approaching a vortex liquid-glass transition and to draw the $H-T$ phase diagram of the flux lines lattice in IBS from a nanoscopic point of view. The mechanism leading to the formation of the Cooper pairs likely involves the spin excitations and $1/T_{1}$ studies allow to follow the evolution of the spin excitations throughout the phase diagram of the IBS and to evidence how the changes in the spin excitations manifest themselves in the superconducting state. More interestingly, several peculiar phenomena characteristic of strongly correlated electron systems have also been evidenced thanks to NMR: the orbital selective behaviour and charge order in the hole-doped IBS, the nematic order and fluctuations extending over a significant region of the phase diagram. Similarly to the cuprates, the presence of several competing energy scales make iron-based superconductors phase diagram rather rich and complex, with several open questions still to be answered, as the mechanisms driving the inhomogeneous charge distribution, the nematic phase and superconductivity. The hunting for those responses will trigger further NMR and NQR research activity in iron-based superconductors in the future.

\acknowledgments

Many of the results presented in this manuscript originate from the collaboration with different colleagues who have deeply collaborated with us in Pavia, in particular Pietro Bonf\`a, Lucia Bossoni, Franziska Hammerath, Matteo Moroni and Samuele Sanna, as well as from the collaboration with colleagues of the broadband NMR community, Giuseppe Allodi, Bernd B\"uchner, Roberto De Renzi, Hajo Grafe, Bill Halperin, Mladen Horvati\'c, Marc-Henri Julien, from the interaction with theorists as Brian Andersen, Massimo Capone, Laura Fanfarillo and Maria Gastiasoro, and with colleagues who provided high quality samples as Sai Aswartham, Paul Canfield, Laura Gozzelino, Yoshiaki Kobayashi, Alberto Martinelli, Andrea Palenzona, Masatoshi Sato, Makariy Tanatar and Sabine Wurmehl. We are very grateful to all of them. Part of the results presented in this work were obtained thanks to the support of MIUR-PRIN2015 Project No. 2015C5SEJJ.

\end{document}